\documentclass[12pt]{article}
\usepackage{subfig}
\usepackage{amssymb,amsmath,euscript}
\usepackage[dvips]{graphicx}
\usepackage{feynmf}
\usepackage{indentfirst}
\usepackage{cite}

\usepackage{geometry}
\begin{document}
\begin{fmffile}{samplepics}

\titlepage
\author{A.V.\,Lipatov, M.A.\,Malyshev, N.P.\,Zotov}
\title{An extended study of prompt photon photoproduction at HERA with $k_T$-factorization}
\maketitle

\begin{center}
{\it D.V.~Skobeltsyn Institute of Nuclear Physics,\\ 
M.V. Lomonosov Moscow State University,
\\119991 Moscow, Russia\/}\\[3mm]
\end{center}

\bigskip

\begin{abstract}
We reconsider prompt photon photoproduction at HERA in the framework 
of the $k_T$-factorization QCD approach. The proposed method is based on the 
${\cal O}(\alpha^2\alpha_s)$ amplitudes 
for $\gamma q\to\gamma gq$ and $\gamma g^*\to\gamma q\bar q$ partonic subprocesses.
Additionally, we take into account the ${\cal O}(\alpha^2\alpha_s^2)$ box 
contributions $\gamma g\to\gamma g$ to the production cross sections. 
The unintegrated (or transverse momentum dependent) parton densities in the proton are determined 
using Kimber-Martin-Ryskin (KMR) prescription. 
Our consideration covers both inclusive and jet associated prompt photon photoproduction rates.
We find that our numerical predictions agree well with the recent data taken by
H1 and ZEUS collaborations at HERA. We demonstrate that the 
box contributions are sizeable and amount up to $\sim15\%$ of the 
calculated total cross section.

\end{abstract}

\vspace{1.8cm}
\section{Introduction}

The prompt (or direct) photon production\footnote{Usually the photons are called "prompt" if they are coupled to the interacting quarks.} at high energies is an important tool to study the hard subprocess dynamics, since the resulting photons are largely insensitive to the effects of the final state hadronization. 
It is a subject of a special interest since, in particular, measurements of total and 
differential cross sections of prompt photons
can be used to constrain the parton densities in a proton.
In the electron-proton collisions at HERA, the ZEUS and H1 Collaborations previously reported data\cite{ZEUS-2000,H1-2004,ZEUS-2006,H1-2009} on the 
production of prompt photons in photoproduction events, where the virtuality of the exchanged 
photon $Q^2$ is lower than 1 GeV$^2$. The next-to-leading order (NLO) 
pQCD calculations\cite{FGH,FH,ZK} underestimate these data by 30---40\%,
especially in rear pseudo-rapidity (electron direction) region,
and the observed disagreement is difficult to explain within conventional theoretical 
uncertainties connected with scale dependence and parametrizations of the parton 
densities \cite{ZEUS-2006,H1-2009}.
However, as it was pointed out in \cite{ZEUS-2006,H1-2009}, the HERA data can be reasonably well described by  calculations \cite{LZ-HERA1,LZ-HERA2}
performed in the framework of the $k_T$-factorization QCD approach\cite{kt1,kt2,kt3,kt4}, which is based on
the Balitsky-Fadin-Kuraev-Lipatov (BFKL)\cite{BFKL1,BFKL2} dynamics of 
parton evolution\footnote{See, for example, reviews \cite{small-x} for introduction to the 
$k_T$-factorization formalism.}.

Very recently new preliminary results for prompt photon photoproduction 
cross sections (both inclusive and associated with the hadronic jet)
have been reported\cite{ZEUS-2013} by the ZEUS Collaboration.
In the present note we apply the $k_T$-factorization approach of QCD 
for the analyses of these data. Note that our previous consideration\cite{LZ-HERA1,LZ-HERA2}
was based on the off-shell ${\cal O}(\alpha^2)$ partonic amplitudes for direct ($\gamma q^*\to\gamma q$) and resolved photon
($q^*g^*\to\gamma q$, $g^*q^*\to\gamma q$ and $q^*\bar q^*\to\gamma g$) contributions
to the photon cross section, where the non-zero transverse momenta of initial 
quarks and gluons were properly taken into account.
Here we reconsider our previous calculations and take into account ${\cal O}(\alpha^2 \alpha_s)$
subprocesses, namely $\gamma g^*\to\gamma q\bar q$ and
$\gamma q\to\gamma gq$.
In the case of photon and associated jet production,
it allows us to take into account the kinematics of the accompanied jet more 
accurately as compared with the previous consideration (see discussion in Section~3).
Additional motivation for our study is that similar consideration\cite{gamma+b}, based 
on the off-shell $2\to 3$ subprocesses, results in a better description
of the Tevatron data on the associated photon and heavy ($b$ or $c$) quark
jet \cite{D0,CDF} as compared to the NLO pQCD predictions \cite{gamma+b-NLO}.
Moreover, we take into account ${\cal O}(\alpha^2 \alpha_s^2)$ contributions from the box 
$\gamma g\to\gamma g$ subprocess\footnote{We will neglect the transverse momenta of 
incoming quarks and gluons in the production amplitudes 
of $\gamma q\to\gamma gq$ and $\gamma g\to\gamma g$ subprocesses,
but keep true off-shell kinematics. See Section 2 for more details.}, which is known to be sizeable\cite{ZK} since the high gluon density region is partially reached.
Following to our previous considerations\cite{LZ-HERA1,LZ-HERA2}, we apply Kimber-Martin-Ryskin (KMR) 
prescription\cite{KMR1,KMR2} to define the unintegrated (or transverse momentum 
dependent) quark and gluon densities in a proton. This approach is a simple 
formalism to construct the unintegrated parton distributions from the known 
collinear ones. We calculate total and
differential inclusive and jet associated prompt photon production cross sections, 
perform a systematic comparison of our predictions with the available H1 and ZEUS data\cite{H1-2009,ZEUS-2013}
and estimate the theoretical uncertainties of our predictions.

The outline of our paper is following. In Section~2 we recall shortly the basic formulas of the $k_T$-factorization approach. In Section~3 we present the numerical results of our calculations and discussion. Section~4 contains our conclusions.

\section{Theoretical framework}

The previous investigation\cite{LZ-HERA1,LZ-HERA2} in the framework of the
$k_T$-factorization approach was based on $\mathcal O(\alpha^2)$ off-shell subprocesses, namely
\begin{gather}
\label{Gq}
\gamma(k_1)+q^*(k_2)\to\gamma(p_1)+q(p_2),\\
\label{qq}
q^*(k_1)+\bar q^*(k_2)\to\gamma(p_1)+g(p_2),\\
\label{qg}
q^*(k_1)+g^*(k_2)\to\gamma(p_1)+q(p_2),\\
\label{gq}
g^*(k_1)+q^*(k_2)\to\gamma(p_1)+q(p_2),
\end{gather}

\noindent
where the subprocess (\ref{Gq}) represents the direct production mechanism, 
and the subprocesses (\ref{qq}) --- (\ref{gq}) represent the
single resolved photon ones, in which the initial photon fluctuated into a hadronic state and a gluon and/or a quark from this hadronic fluctuation takes part in the hard 
interaction\footnote{The subprocesses (2) --- (4), which are formally of $\mathcal O(\alpha \alpha_s)$,
give also the $\mathcal O(\alpha^2)$ contributions since the parton densities in a photon at leading-order have a behavior proportional to $\alpha \ln\mu^2/\Lambda_{\rm QCD}^2 \sim {\alpha/\alpha_s}$.}.
In the present note we concentrate on $\mathcal O(\alpha^2\alpha_s)$ subproccesses:
\begin{gather}
\label{Gq1}
\gamma(k_1)+q(k_2)\to\gamma(p_1)+g(p_2)+q(p_3),\\
\label{Gg}
\gamma(k_1)+g^*(k_2)\to\gamma(p_1)+q(p_2)+\bar q(p_3),
\end{gather}

\noindent 
where the relevant four-momenta are given in the parentheses. 
Additionally, for the first time
in the framework of $k_T$-factorization, we take into account
box contribution:
\begin{equation}
\label{box}
\gamma(k_1)+g(k_2)\to\gamma(p_1)+g(p_2).
\end{equation}

\begin{figure}
\center
\begin{fmfgraph*}(120,60)
\fmfleft{i1,i2}
\fmfright{o1,o2}
\fmf{gluon}{i1,v1}
\fmf{fermion,tension=0.5}{v1,v2}
\fmf{gluon}{v2,o1}
\fmf{boson}{i2,v3}
\fmf{fermion,tension=0.5}{v4,v3}
\fmf{boson}{v4,o2}
%\fmf{fermion}{o2,v4,v3,i2}
\fmf{fermion,tension=0.4}{v3,v1}
\fmf{fermion,tension=0.4}{v2,v4}
\end{fmfgraph*}
\hspace*{1cm}
\begin{fmfgraph*}(120,60)
\fmfleft{i1,i2}
\fmfright{o1,o2}
\fmf{gluon}{i1,v1}
\fmf{fermion,tension=0.5}{v1,v2}
\fmf{boson}{v2,o1}
\fmf{boson}{i2,v3}
\fmf{fermion,tension=0.5}{v4,v3}
\fmf{gluon}{v4,o2}
%\fmf{fermion}{o2,v4,v3,i2}
\fmf{fermion,tension=0.4}{v3,v1}
\fmf{fermion,tension=0.4}{v2,v4}
\end{fmfgraph*}
\hspace*{1cm}
\begin{fmfgraph*}(120,60)
\fmfleft{i1,i2}
\fmfright{o1,o2}
\fmf{gluon}{i1,v1}
\fmf{fermion,tension=0.5}{v1,v2}
\fmf{gluon}{v2,o1}
\fmf{boson}{i2,v3}
\fmf{fermion,tension=0.5}{v3,v4}
\fmf{boson}{v4,o2}
%\fmf{fermion}{o2,v4,v3,i2}
\fmf{plain,tension=0.4}{v2,v3}
\fmf{fermion,tension=0.4}{v4,v1}
\end{fmfgraph*}
\caption{\it The 'box'-diagrams. One should also take into account diagrams with the opposite direction of the fermion loop.}
\label{box-graph}
\end{figure}
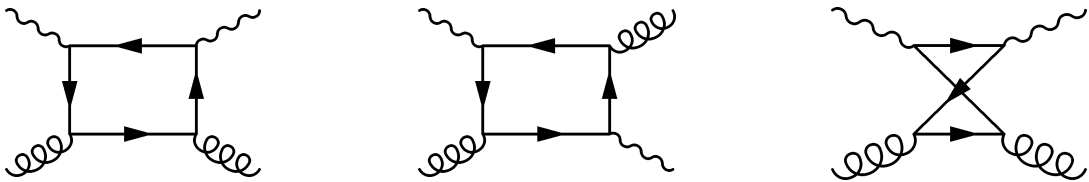

\noindent 
This subprocess, formally is of $\mathcal O(\alpha^2\alpha_s^2)$,
is known to be sizeable\cite{ZK} due to high gluon luminosity in
the probed kinematical region and it was taken into account in the standard QCD calcualtions \cite{FH}. 
The corresponding Feynman diagrams for the subprocesses under consideration 
are shown on Figs.~\ref{2to3} and~\ref{box-graph}.
Note that, in contrast with the collinear QCD approximation, the subprocesses (\ref{Gq}) --- (\ref{qg}) are effectively included in (\ref{Gq1}) and (\ref{Gg}) 
in the $k_T$-factorization approach, and only subprocess (\ref{gq}) stays 
out. However, according to the estimates in \cite{LZ-HERA1}, 
this mechanism gives only a few percent contribution to the cross section in the 
kinematical region covered by the H1 and ZEUS experiments, 
so we can neglect it.
We neglect also the contributions from the so-called fragmentation mechanisms 
since the the isolation cut application \cite{H1-2009,ZEUS-2013} reduces these contributions to less than 10\% 
of the visible cross section. Note that the isolation cuts and additional conditions which preserve our 
calculations from divergences were specially discussed in \cite{LZ-HERA2}.

The amplitudes for the subprocesses (\ref{Gq1}) and (\ref{Gg}) can be written as follows:
\begin{gather}
\mathcal M(\gamma q\to\gamma gq)=e^2 e_q^2\,g\, t^a \epsilon_\lambda(k_1)\epsilon_\mu^*(p_1)\epsilon_\nu^*(p_2)\sum_{i=1}^{6}\mathcal F_i^{\mu\nu\lambda},\\
\mathcal M(\gamma g^*\to\gamma q\bar q)= e^2 e_q^2\,g\,t^a\epsilon_\mu(k_1)\epsilon_\nu(k_2)\epsilon_\lambda^*(p_1)\sum_{i=1}^{6}\mathcal G_i^{\mu\nu\lambda},
\end{gather}

\noindent where
\begin{gather}
\mathcal F_1^{\mu\nu\lambda}=\bar u(p_3)\gamma^\nu\frac{\hat p_2+\hat p_3+m}{(p_2+p_3)^2-m^2}\gamma^\mu\frac{\hat k_1+\hat k_2+m}{(k_1+k_2)^2-m^2}\gamma^\lambda u(k_2),\\
\mathcal F_2^{\mu\nu\lambda}=\bar u(p_3)\gamma^\mu\frac{\hat p_1+\hat p_3+m}{(p_1+p_3)^2-m^2}\gamma^\nu\frac{\hat k_1+\hat k_2+m}{(k_1+k_2)^2-m^2}\gamma^\lambda u(k_2),\\
\mathcal F_3^{\mu\nu\lambda}(=\bar u(p_3)\gamma^\nu\frac{\hat p_2+\hat p_3+m}{(p_2+p_3)^2-m^2}\gamma^\lambda\frac{\hat k_2-\hat p_1+m}{(k_2-p_1)^2-m^2}\gamma^\mu u(k_2),\\
\mathcal F_4^{\mu\nu\lambda}=\bar u(p_3)\gamma^\lambda\frac{\hat p_3-\hat k_1+m}{(p_3-k_1)^2-m^2}\gamma^\mu\frac{\hat k_2-\hat p_2+m}{(k_2-p_2)^2-m^2}\gamma^\nu u(k_2),\\
\mathcal F_5^{\mu\nu\lambda}(=\bar u(p_3)\gamma^\mu\frac{\hat p_1+\hat p_3+m}{(p_1+p_3)^2-m^2}\gamma^\lambda\frac{\hat k_2-\hat p_2+m}{(k_2-p_2)^2-m^2}\gamma^\nu u(k_2),\\
\mathcal F_6^{\mu\nu\lambda}=\bar u(p_3)\gamma^\lambda\frac{\hat p_3-\hat k_1+m}{(p_3-k_1)^2-m^2}\gamma^\nu\frac{\hat k_2-\hat p_3+m}{(k_2-p_3)^2-m^2}\gamma^\mu u(k_2),
\end{gather}

\noindent and
\begin{gather}
\mathcal G_1^{\mu\nu\lambda}=\bar u(p_2)\gamma^\lambda\frac{\hat p_2-\hat p_1+m}{(p_2-p_1)^2-m^2}\gamma^\mu\frac{\hat k_2-\hat p_3+m}{(k_2-p_3)^2-m^2}\gamma^\nu u(p_3),\\
\mathcal G_2^{\mu\nu\lambda}=\bar u(p_2)\gamma^\mu\frac{\hat p_2-\hat k_1+m}{(p_2-k_1)^2-m^2}\gamma^\lambda\frac{\hat k_2-\hat p_3+m}{(k_2-p_3)^2-m^2}\gamma^\nu u(p_3),\\
\mathcal G_3^{\mu\nu\lambda}=\bar u(p_2)\gamma^\mu\frac{\hat k_1-\hat p_2-m}{(k_1-p_2)^2-m^2}\gamma^\nu\frac{\hat p_3+\hat p_1-m}{(p_3+p_1)^2-m^2}\gamma^\lambda u(p_3),\\
\mathcal G_4^{\mu\nu\lambda}=\bar u(p_2)\gamma^\lambda\frac{\hat p_2-\hat p_1+m}{(p_2-p_1)^2-m^2}\gamma^\nu\frac{\hat k_1-\hat p_3+m}{(k_1-p_3)^2-m^2}\gamma^\mu u(p_3),\\
\mathcal G_5^{\mu\nu\lambda}=\bar u(p_2)\gamma^\nu\frac{\hat p_2-\hat k_2+m}{(p_2-k_2)^2-m^2}\gamma^\lambda\frac{\hat k_1-\hat p_3+m}{(k_1-p_3)^2-m^2}\gamma^\mu u(p_3),\\
\mathcal G_6^{\mu\nu\lambda}=\bar u(p_2)\gamma^\nu\frac{\hat k_2-\hat p_2-m}{(k_2-p_2)^2-m^2}\gamma^\mu\frac{\hat p_3+\hat p_1-m}{(p_3+p_1)^2-m^2}\gamma^\lambda u(p_3).
\end{gather}

\noindent
Here $e$ is the electric charge, $e_q$ is the (fractional) charge of quark having mass $m$, 
$g$ is the strong charge, $\epsilon(k_1)$, $\epsilon(k_2)$, $\epsilon(p_1)$ and $\epsilon(p_2)$ are the polarization four-vectors of the corresponding particles and $a$ is the eight-fold color index.
Note again that in (8) we neglected the transverse momentum of initial quark.
The calculation of the off-shell matrix elements listed above is straightforward. Here
we would like to only mention that, according to the $k_T$-factorization
prescription \cite{kt2,kt3,kt4,BFKL1}, the summation over the incoming off-shell gluon polarizations is carried out
with
\begin{equation}
\label{sumgluon}
\sum \epsilon^\mu\epsilon^{*\nu}=\frac{k^\mu_T k^\nu_T}{\mathbf k^2_T},
\end{equation}

\noindent
where $k_T$ is the gluon non-zero transverse momentum. For the photons 
and outgoing on-shell gluon the summation over their polarizations 
can be performed with the usual covariant formula:
\begin{equation}
\sum \epsilon^\mu\epsilon^{*\nu} = -g^{\mu\nu}.
\end{equation}

\noindent In other respects we follow the standard QCD Feynman rules.
The evaluation of the emerging traces was done using the algebraic manipulation 
system \textsc{form} \cite{FORM}.
Finally, concerning the box contribution (7), the corresponding amplitude squared was calculated a long
time ago in the on-shell limit ${\mathbf k}_{2T}^2 \to 0$. 
The simple analytical expression can be found, for example, in\cite{box}. 
In our phenomenological study, we apply this expression,
however we keep the exact off-shell kinematics (see \cite{diphoton} for more details).

According to the $k_T$-factorization prescription, to calculate the cross 
section one should convolute off-shell partonic cross sections
with the relevant unintegrated (transverse momentum dependent) 
quark and/or gluon distributions in a proton.
In the present paper we use the KMR approximation\cite{KMR1,KMR2} to determine it. 
The KMR approach is a formalism to construct the unintegrated parton densities
$f_a(x,{\mathbf k}_T^2,\mu^2)$ from the known collinear parton
distributions $xa(x,\mu^2)$, where $a = g$ or $a = q$. 
In this approximation, the unintegrated quark and 
gluon densities are given by\cite{KMR1,KMR2}
\begin{gather}
\label{fq}
  \displaystyle f_q(x,{\mathbf k}_T^2,\mu^2) = T_q({\mathbf k}_T^2,\mu^2) {\alpha_s({\mathbf k}_T^2)\over 2\pi} \times \atop {
  \displaystyle \times \int\limits_x^1 dz \left[P_{qq}(z) {x\over z} q\left({x\over z},{\mathbf k}_T^2\right) \Theta\left(\Delta - z\right) + P_{qg}(z) {x\over z} g\left({x\over z},{\mathbf k}_T^2\right) \right],} 
\\
\label{fg}
  \displaystyle f_g(x,{\mathbf k}_T^2,\mu^2) = T_g({\mathbf k}_T^2,\mu^2) {\alpha_s({\mathbf k}_T^2)\over 2\pi} \times \atop {
  \displaystyle \times \int\limits_x^1 dz \left[\sum_q P_{gq}(z) {x\over z} q\left({x\over z},{\mathbf k}_T^2\right) + P_{gg}(z) {x\over z} g\left({x\over z},{\mathbf k}_T^2\right)\Theta\left(\Delta - z\right) \right],} 
\end{gather}

\noindent
where $P_{ab}(z)$ are the usual unregulated LO DGLAP splitting 
functions. The theta functions which appear 
in~(\ref{fq}) and~(\ref{fg}) imply the angular-ordering constraint $\Delta = \mu/(\mu + |{\mathbf k}_T|)$ 
specifically to the last evolution step to regulate the soft gluon
singularities.
The Sudakov form factors $T_q({\mathbf k}_T^2,\mu^2)$ and 
$T_g({\mathbf k}_T^2,\mu^2)$ which appear in (\ref{fq}) and (\ref{fg}) enable us to include logarithmic loop corrections
to the calculated cross sections\footnote{Numerically, in (\ref{fq}) and (\ref{fg}) we applied the 
MSTW'2008 parton distributions \cite{MSTW}.}. 

The contributions to the prompt photon cross sections from
subprocesses (5) --- (7) can be easily written as follows:
\begin{equation}
\displaystyle \sigma(\gamma p\to\gamma X)=\sum_q\int\frac{E_T^\gamma}{128\pi^3\, y (x_2 s)^2}\overline{|\mathcal M(\gamma q\to\gamma qg)|^2} f_{q}(x_2,\mathbf k_{2T}^2,\mu^2) \times \atop {
\displaystyle \times dy_1\,dy_2\,dy^\gamma\,dE_T^\gamma\,d\mathbf k_{2T}^2\frac{d\phi_2}{2\pi}\frac{d\psi_1}{2\pi}\frac{d\psi_2}{2\pi}\frac{d\psi^\gamma}{2\pi}},
\label{sigma1}
\end{equation}
\begin{equation}
\displaystyle \sigma(\gamma p\to\gamma X)=\int\frac{E_T^\gamma}{128\pi^3\,y(x_2 s)^2}\overline{|\mathcal M(\gamma g^*\to\gamma qq)|^2} f_g(x_2,\mathbf k_{2T}^2,\mu^2) \times \atop {
\displaystyle \times dy_1\,dy_2\,dy^\gamma\,dE_T^\gamma\,d\mathbf k_{2T}^2\frac{d\phi_2}{2\pi}\frac{d\psi_1}{2\pi}\frac{d\psi_2}{2\pi}\frac{d\psi^\gamma}{2\pi}},
\label{sigma2}
\end{equation}
\begin{equation}
\displaystyle \sigma(\gamma p\to\gamma X)=\int\frac{E_T^\gamma}{8\pi\,y(x_2 s)^2}\overline{|\mathcal M(\gamma g \to\gamma g)|^2} f_g(x_2,\mathbf k_{2T}^2,\mu^2) \times \atop {
\displaystyle \times dy^\gamma\,dE_T^\gamma\,d\mathbf k_{2T}^2\frac{d\phi_2}{2\pi}\frac{d\psi^\gamma}{2\pi}},
\label{sigma3}
\end{equation}

\noindent
where $s$ is the total center-of-mass energy of the collision, 
$\phi_2$ is the azimuthal angle\footnote{The angle of the transverse momentum 
${\mathbf k}_T$ in the plane perpendicular to $OZ$ axis.} 
of initial quark or gluon having fraction $x_2$ of the initial proton longitudinal momentum and 
non-zero transverse momentum $|\mathbf k_{2T}|\ne0$, $E_T^\gamma$, $y^\gamma$, $\phi^\gamma$ 
are the transverse energy, rapidity and azimuthal angle of the produced photon, 
$y_1$, $y_2$, $\psi_1$ and $\psi_2$ are the rapidities and azimuthal angles of the outcoming partons,
respectively.

The experimental data\cite{H1-2009,ZEUS-2013} refer to the prompt photon production in $ep$ collisions, where the electron emits a quasi-real ($Q^2\sim0$) photon. Thus $\gamma p$ cross sections (26) --- (28) should be weighted with the photon flux in the electron:
\begin{equation}
d\sigma(ep\to e'+\gamma+X)=\int f_{\gamma/e}(y)d\sigma(\gamma p\to\gamma+X)dy,
\end{equation}

\noindent
where $y$ is the fraction of the initial electron energy carried by the photon in the laboratory frame. We use here the Weizacker-Williams approximation for the bremsstrahlung photon distribution from the electron:
\begin{equation}
f_{\gamma/e}(y)=\frac{\alpha}{2\pi}\left(\frac{1+(1-y)^2}{y}\ln\frac{Q^2_{\text{max}}}{Q^2_{\text{min}}}+2m^2_ey\left(\frac{1}{Q^2_{\text{max}}}-\frac{1}{Q^2_{\text{min}}}\right)\right),
\end{equation}

\noindent
where $m_e$ is the electron mass, $Q^2_{\text{min}}=m_e^2y^2/(1-y)^2$ and $Q^2_{\text{max}}=1$~GeV$^2$, which is a typical value for the recent photoproduction measurements at the HERA collider.

The multidimensional integration in (26) --- (28) has been performed by means of the Monte-Carlo technique, using the routine \textsc{vegas}\cite{VEGAS}. The full C++ code is available from the authors on request\footnote{malyshev@theory.sinp.msu.ru}.

\section{Numerical results}
\label{Results}

Now we are in a position to present our numerical results. First we describe our theoretical input and the kinematical conditions. 
The renormalization and factorization scales were taken as $\mu_R=\mu_F=\xi E^\gamma_T$, where 
the parameter $\xi$ was varied between 1/2 and 2 about the default value $\xi=1$ to estimate the scale uncertainties of our calculations. We neglected quark masses and used the standard LO formula for the strong coupling constant $\alpha_s(\mu^2)$ with $n_f=4$ massless quark flavours and $\Lambda_{\rm QCD}=200$ MeV, such that $\alpha_s(M_Z^2)=0.1232$.

The experimental data for the inclusive prompt photon photoproduction at HERA 
were taken by both the H1 and ZEUS collaborations. The H1 data\cite{H1-2009} were obtained in the following kinematical region\footnote{Here and in the following all kinematic quantities are given in the laboratory frame with positive Z axis direction given by the proton beam.}: $6<E_T^\gamma<15$ GeV and $-1.0<\eta^\gamma<2.4$. The fraction $y$ of the electron energy  transferred to the photon is restricted to the range $0.1<y<0.7$. Very recent ZEUS measurements\cite{ZEUS-2013} refer to the region defined by $6<E_T^\gamma<15$ GeV, $-0.7<\eta^\gamma<0.9$ and $0.2<y<0.7$.
In the case of jet associated prompt photon photoproduction the restrictions on the prompt photon transverse momentum and pseudo-rapidity 
are the same as in the inclusive production 
case. For the jets, the cuts which were applied in the H1 and ZEUS analyses are $E_T^{\rm jet}>4.5$ GeV,  $-1.3<\eta^{\rm jet}<2.3$ and $4<E_T^{\rm jet}<35$ GeV and $-1.5<\eta^{\rm jet}<1.8$, respectively.
The data\cite{H1-2009,ZEUS-2013} were obtained with the electron energy $E_e=27.6$ GeV and the proton energy $E_p=920$ GeV.

\begin{figure}[t]
\centering
\subfloat[]%optionally add here a short text as a label]
{\label{fig:image_11a}
\includegraphics[width=0.5\textwidth]{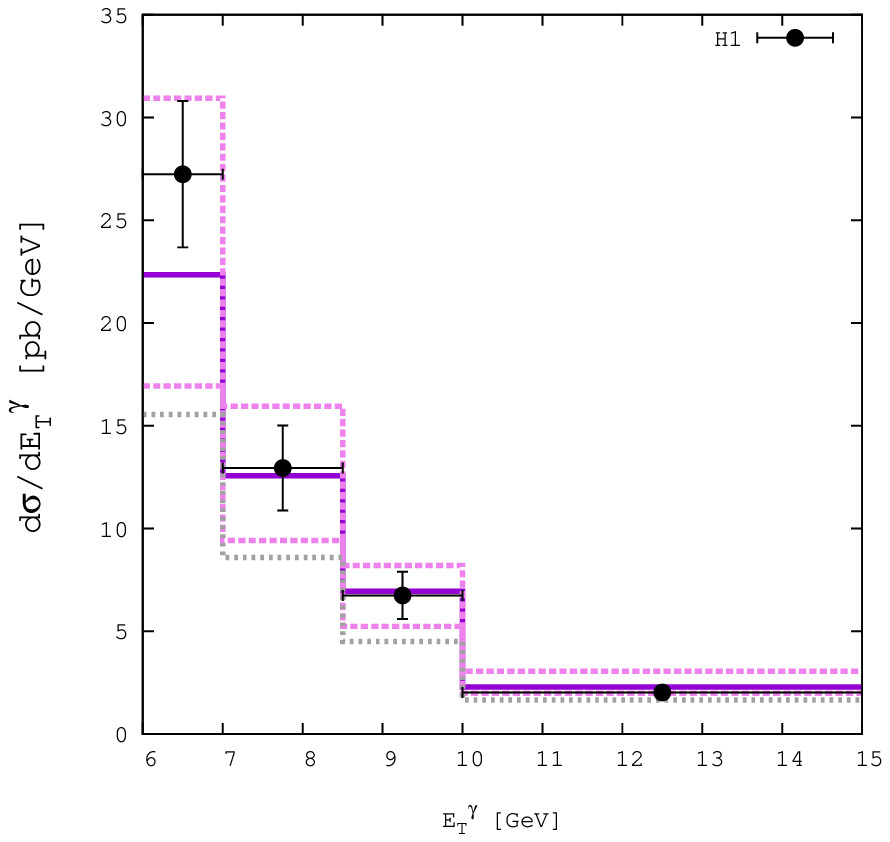}}
\subfloat[]
{\label{fig:image_21a}
\includegraphics[width=0.5\textwidth]{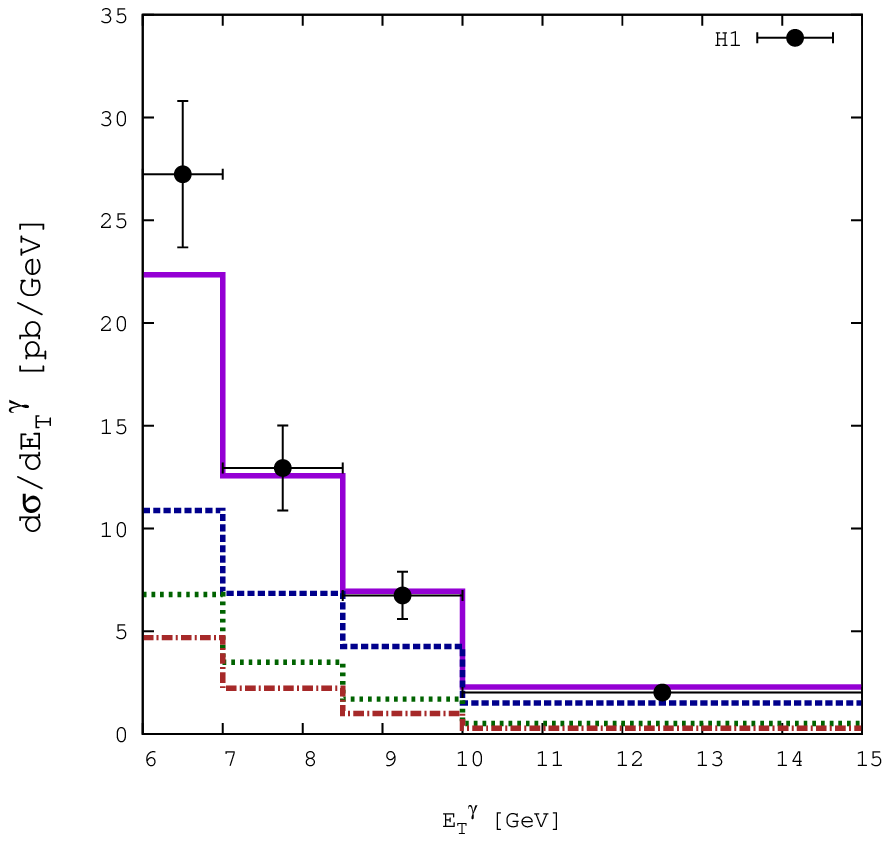}}

\subfloat[]{
\label{fig:image_31a}
\includegraphics[width=0.5\textwidth]{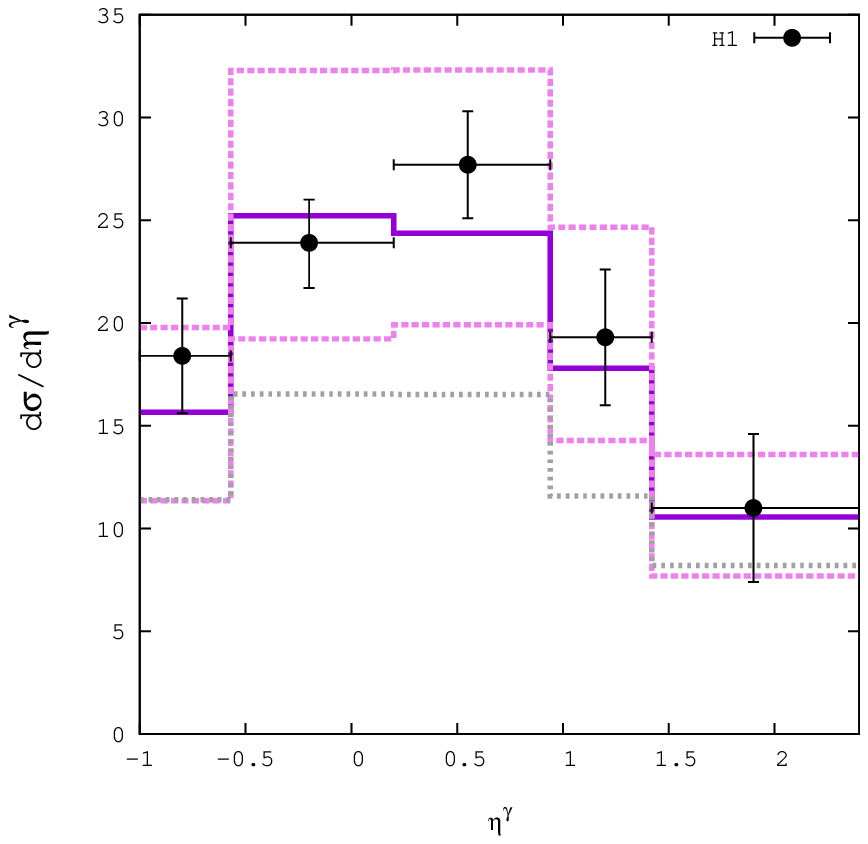}}
\subfloat[]{
\label{fig:image_41a}
\includegraphics[width=0.5\textwidth]{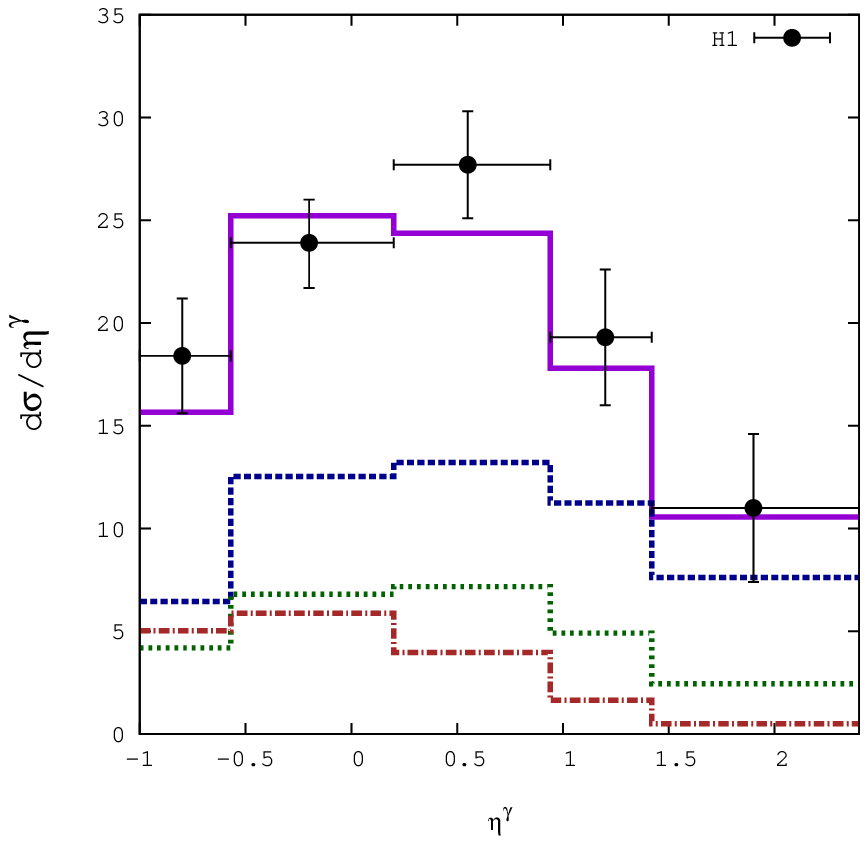}}

\centering
\caption{\label{inclusive-H1} \it{The inclusive prompt photon photoproduction cross
section as a function of photon transverse energy $E_T^\gamma$ and pseudo-rapidity
$\eta^\gamma$ at HERA. The left panel: the solid curve corresponds to the KMR
predictions at the default scale $\mu=E_T^\gamma$ , whereas the upper and lower
dashed curves correspond to scale variations described in the text; the dotted line
represents the results obtained in previous papers
\cite{LZ-HERA1,LZ-HERA2}. The right panel: the solid curve represents the total
cross section; dashed, dotted and dash-dotted lines correspond to the contributions
from $\gamma q\to\gamma gq$, $\gamma g\to\gamma q\bar q$ and $\gamma g\to\gamma g$
respectively. The experimental data are from H1 \cite{H1-2009}.
}}
\end{figure}

\begin{figure}[t]
\centering
\subfloat[]%optionally add here a short text as a label]
{\label{fig:image_11b}
\includegraphics[width=0.5\textwidth]{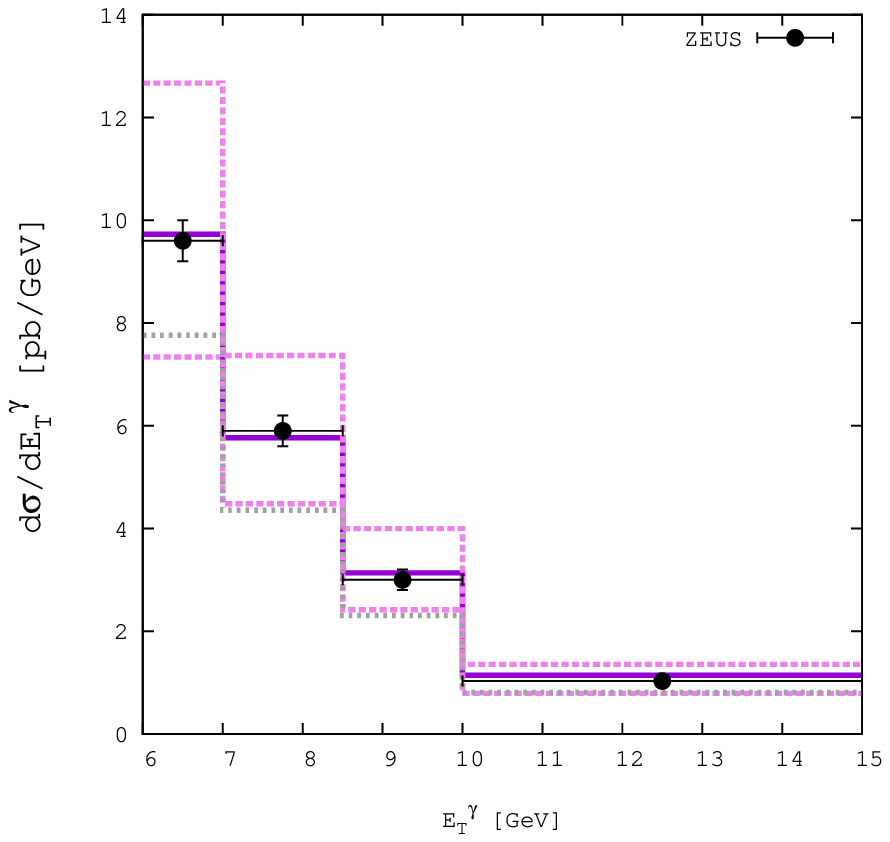}}
\subfloat[]
{\label{fig:image_21b}
\includegraphics[width=0.5\textwidth]{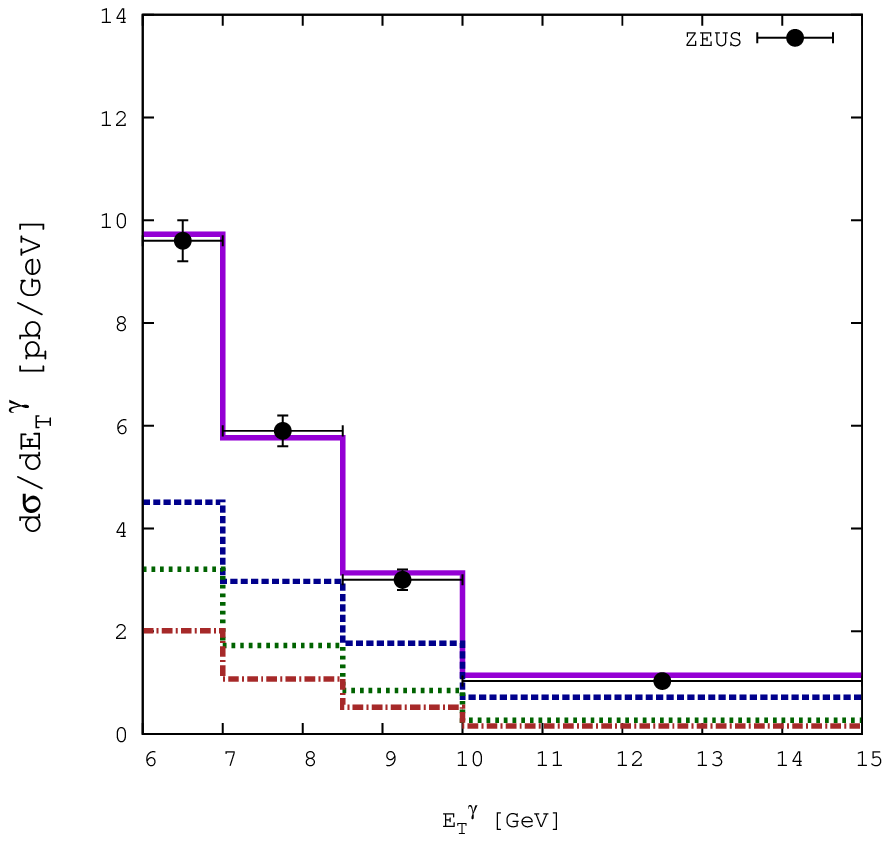}}

\subfloat[]{
\label{fig:image_31b}
\includegraphics[width=0.5\textwidth]{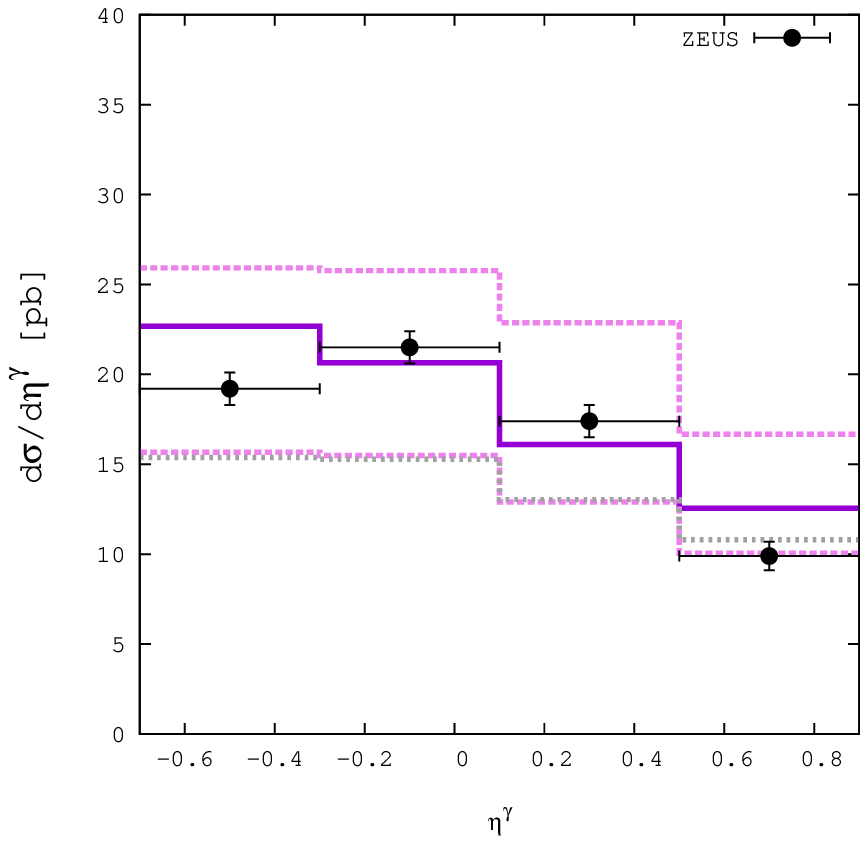}}
\subfloat[]{
\label{fig:image_41b}
\includegraphics[width=0.5\textwidth]{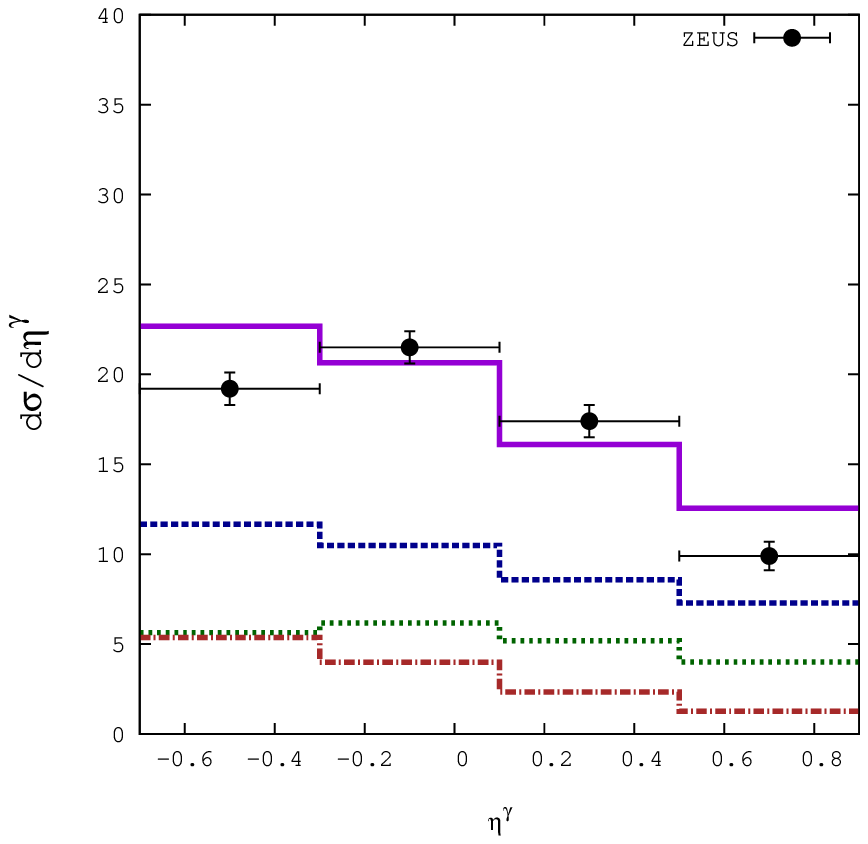}}

\centering
\caption{\label{inclusive-ZEUS} \it{The inclusive prompt photon photoproduction
cross section as a function of photon transverse energy $E_T^\gamma$ and
pseudo-rapidity $\eta^\gamma$ at HERA. The notations of the histograms are the same as in Fig.~\ref{inclusive-H1}. The experimental data are from ZEUS \cite{ZEUS-2013}.
}}
\end{figure}

The transverse energy and pseudo-rapidity distributions of the inclusive prompt photon
production are shown in Figs.~\ref{inclusive-H1} --- \ref{inclusive-ZEUS} in
comparison with the H1 and ZEUS data\cite{H1-2009,ZEUS-2013}. 
In the left panels, the solid histograms correspond to the predictions obtained
from (26) --- (28) at the default scale. The dashed histograms represent the corresponding 
theoretical uncertainties estimated as it was described above. We find that our predictions reasonably well describe a full set of the available experimental data. Moreover, the shape and absolute normalization of the measured 
cross sections are adequately reproduced within the theoretical and experimental uncertainties.
Additionally we plot predictions based on the $2\to2$ subprocesses (1) --- (4), 
as it was done in our previous paper\cite{LZ-HERA1,LZ-HERA2} (dotted histograms in the left panels)\footnote{The depicted $2\to2$ subprocesses based results differ a little from the ones presented in \cite{LZ-HERA1,LZ-HERA2}, since the former have been obtained with the MSTW parton distributions instead of older GRV94 set as the input for KMR.}.
One can see some enhancement of the calculated cross sections due
to, in particular, the box subprocess (7) included into our present consideration.
The relative contributions of different subprocesses to the prompt photon cross section
are shown on the right panels of Figs.~\ref{inclusive-H1} --- \ref{inclusive-ZEUS}. We find that while the subprocess (\ref{Gq1}) 
dominates, the box subprocess (\ref{box}) contributes significantly to the predicted 
cross section, specially at negative photon pseudorapidities. In this region,
the box contribution is comparable with the contribution from the subprocess (\ref{Gg}),
and it amounts up to $\sim15\%$ of the total cross section of inclusive prompt photon production.

Now we turn to the prompt photon production associated with the hadronic jet. 
In the previous consideration\cite{LZ-HERA1,LZ-HERA2}, to calculate the semi-inclusive production 
rates some approximation was applied. So, it was noted that the 
produced photon is accompanied by a number of partons radiated in the course 
of the parton evolution. On the average, the parton transverse momentum decreases from the hard interaction box towards the proton. As an approximation, it was assumed 
that the parton $k^\prime$, emitted in the last
evolution step, compensates the whole transverse momentum of the parton participating in
the hard subprocess, i.e. ${\mathbf{k}_T^\prime} \simeq - {\mathbf{k}_T}$. 
All the other emitted partons are collected together
in the proton remnant, which is assumed to carry only a negligible transverse momentum
compared to ${\mathbf{k}_T^\prime}$. This parton gives rise to a final hadron jet 
with $E_T^{\rm jet} = |{\mathbf{k}_T^\prime}|$ in addition to
the jet produced in the hard subprocess. From these hadron jets the one carrying
the largest transverse energy has been choosen \cite{LZ-HERA1,LZ-HERA2}.
In the present paper we use the same approximation. However, since we use $2\to 3$ rather than 
$2\to 2$ subprocesses, the jet production kinematics is described more accurately than 
it was done previously, because of the production of jets mainly comes from the hard subprocesses.

\begin{figure}[t]
\centering
\subfloat[]%optionally add here a short text as a label]
{\label{fig:image_12a1}
\includegraphics[width=0.5\textwidth]{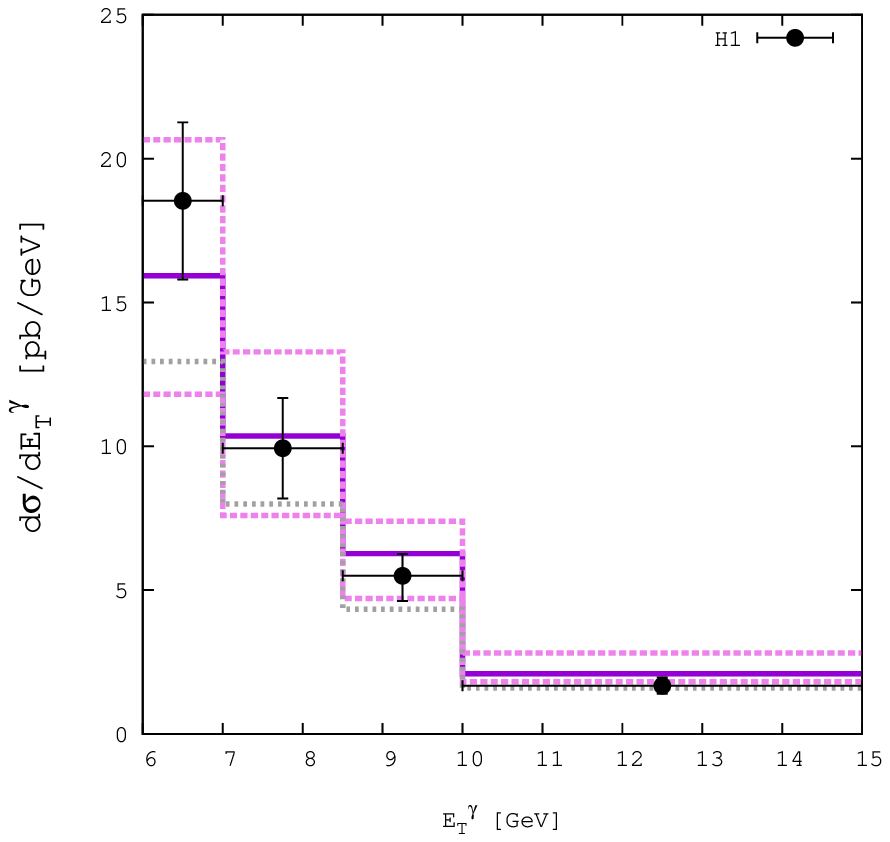}}
\subfloat[]
{\label{fig:image_22a1}
\includegraphics[width=0.5\textwidth]{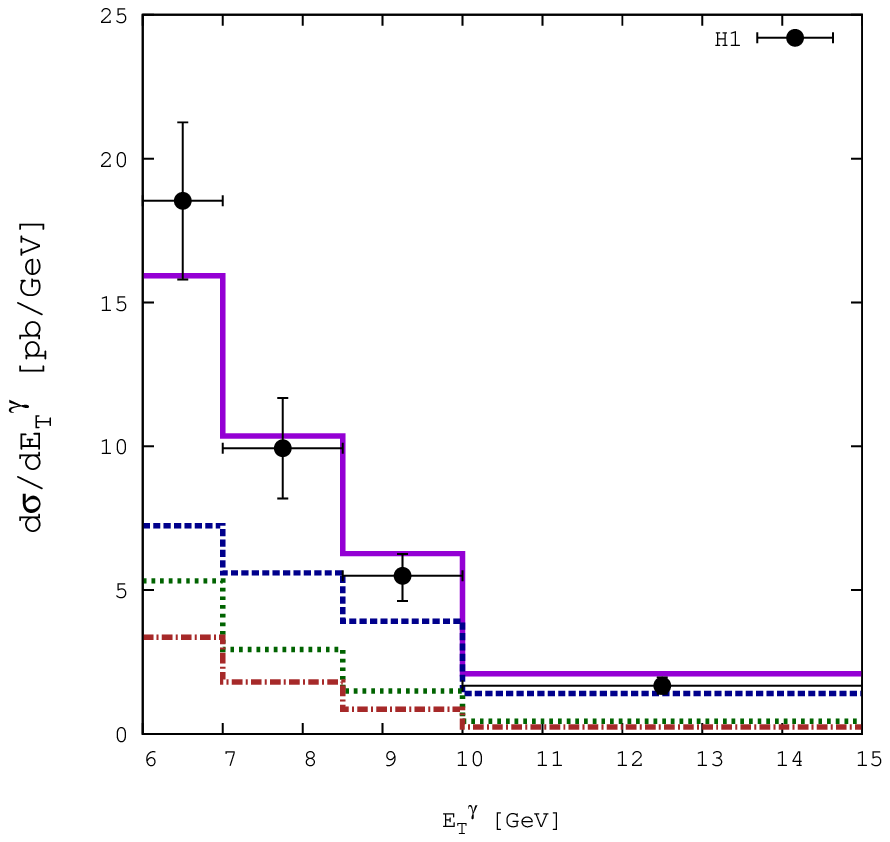}}

\subfloat[]%optionally add here a short text as a label]
{\label{fig:image_12b1}
\includegraphics[width=0.5\textwidth]{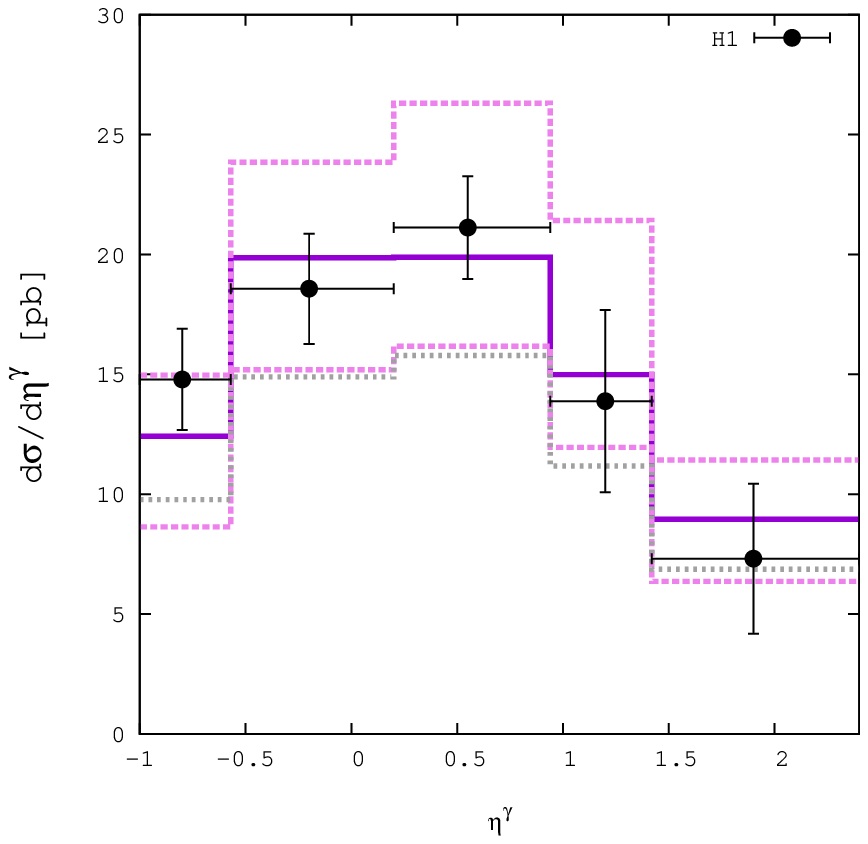}}
\subfloat[]
{\label{fig:image_22b1}
\includegraphics[width=0.5\textwidth]{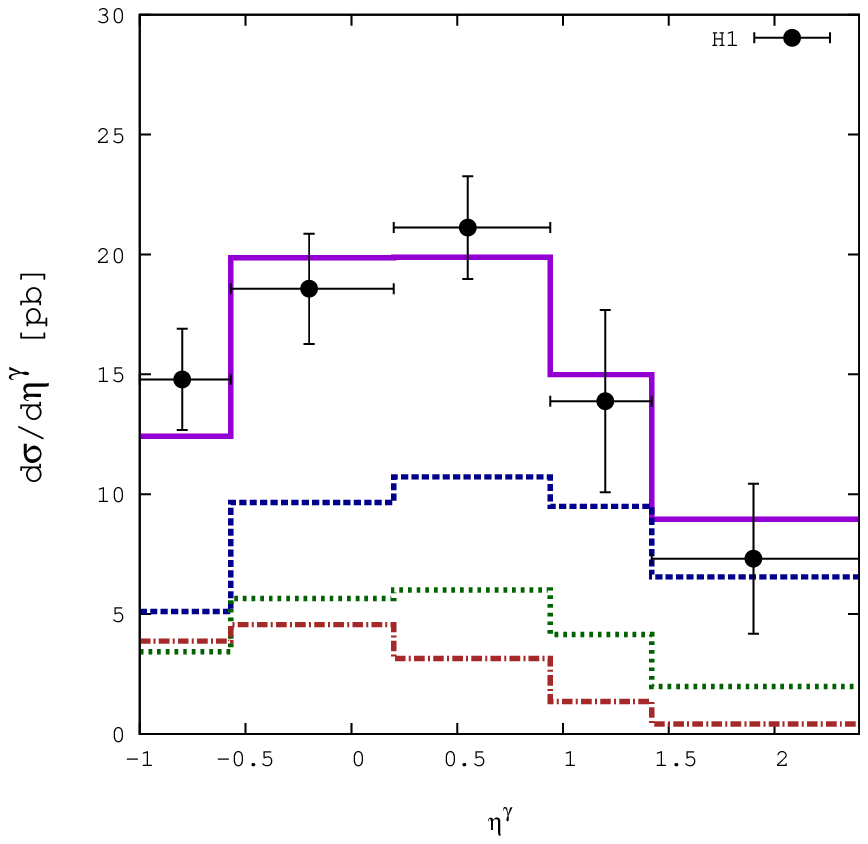}}

\centering
\caption{\label{H11} \it{The associated with a jet prompt photon photoproduction
cross section as a function of photon transverse energy $E_T^\gamma$ and
pseudo-rapidity $\eta^\gamma$ at HERA. The notations of the histograms are the same as in Fig.~\ref{inclusive-H1}. The experimental data are from H1 \cite{H1-2009}.
}}
\end{figure}

\begin{figure}[t]
\centering
\subfloat[]%optionally add here a short text as a label]
{\label{fig:image_12c1}
\includegraphics[width=0.5\textwidth]{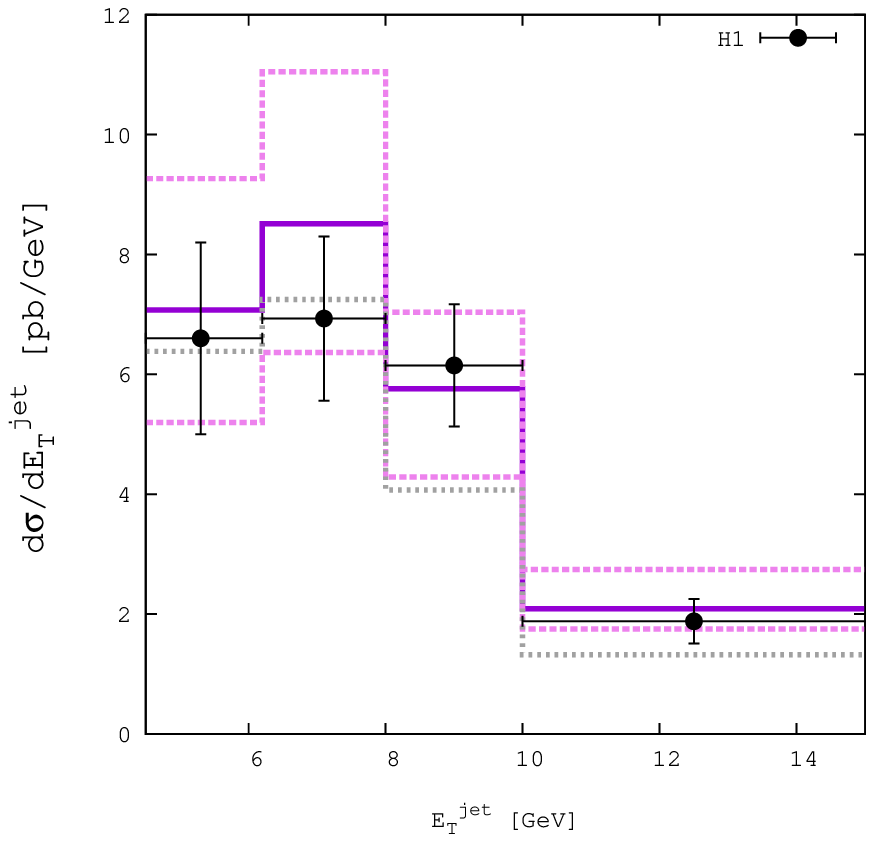}}
\subfloat[]
{\label{fig:image_22c1}
\includegraphics[width=0.5\textwidth]{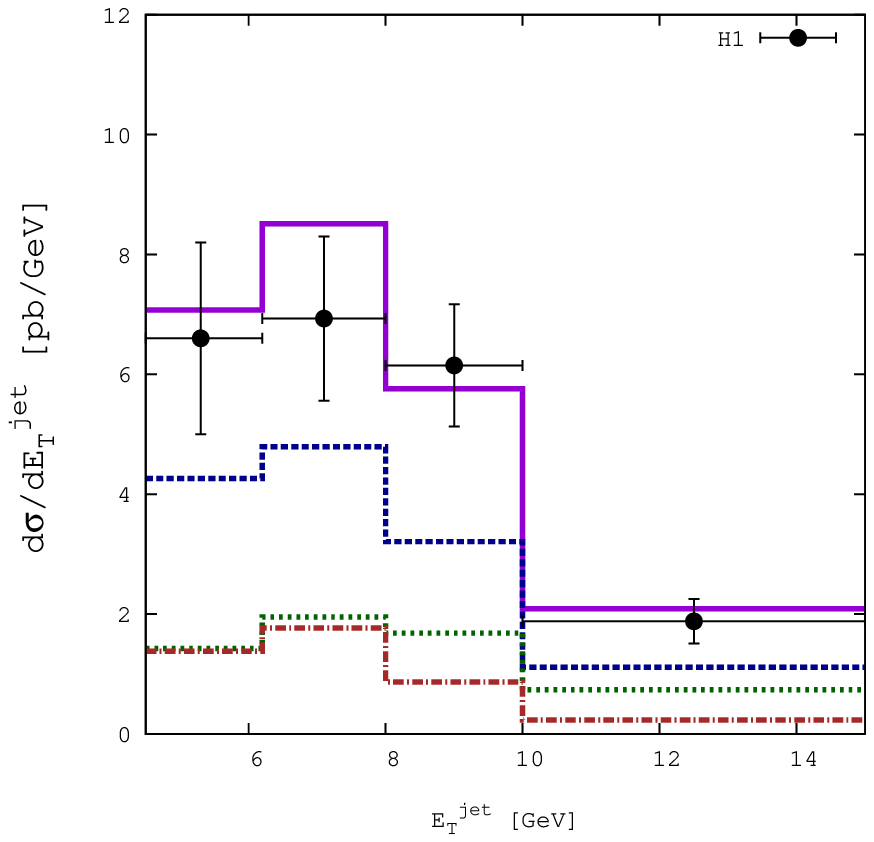}}

\subfloat[]{
\label{fig:image_32d1}
\includegraphics[width=0.5\textwidth]{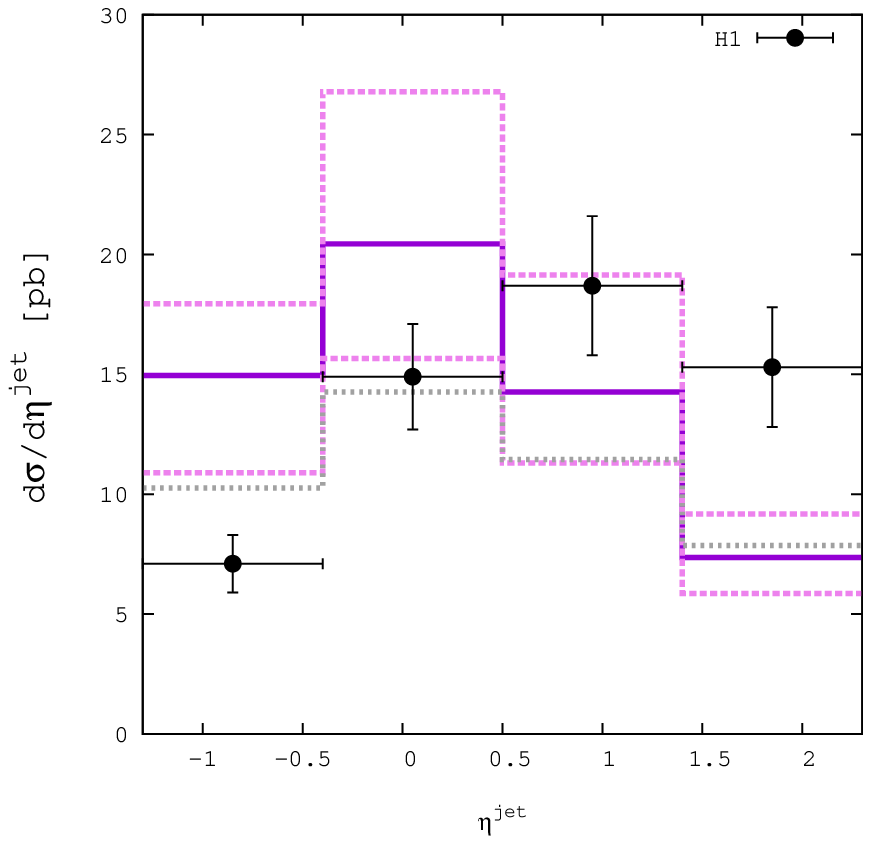}}
\subfloat[]{
\label{fig:image_42d1}
\includegraphics[width=0.5\textwidth]{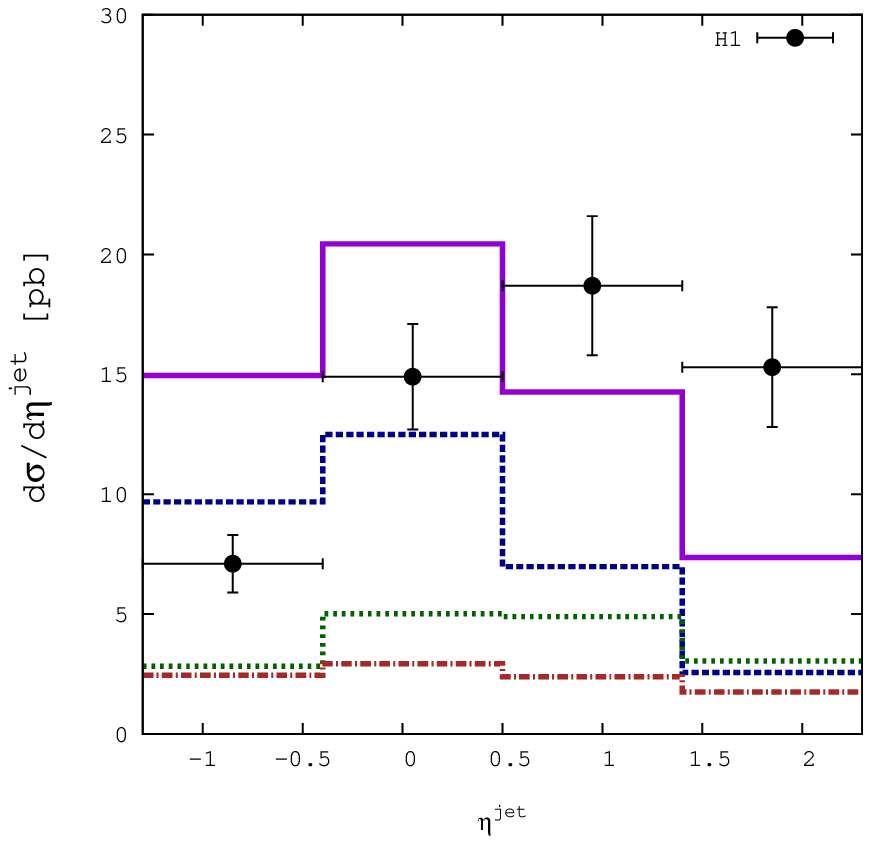}}

\centering
\caption{\label{H12} \it{The associated with a jet prompt photon photoproduction
cross section as a function of jet transverse energies $E_T^{jet}$ and
pseudo-rapidities $\eta^{jet}$ at HERA. The notations of the histograms are the same as in Fig.~\ref{inclusive-H1}. The experimental data are from H1 \cite{H1-2009}.
}}
\end{figure}

\begin{figure}[t]
\centering
\subfloat[]%optionally add here a short text as a label]
{\label{fig:image_12a2}
\includegraphics[width=0.5\textwidth]{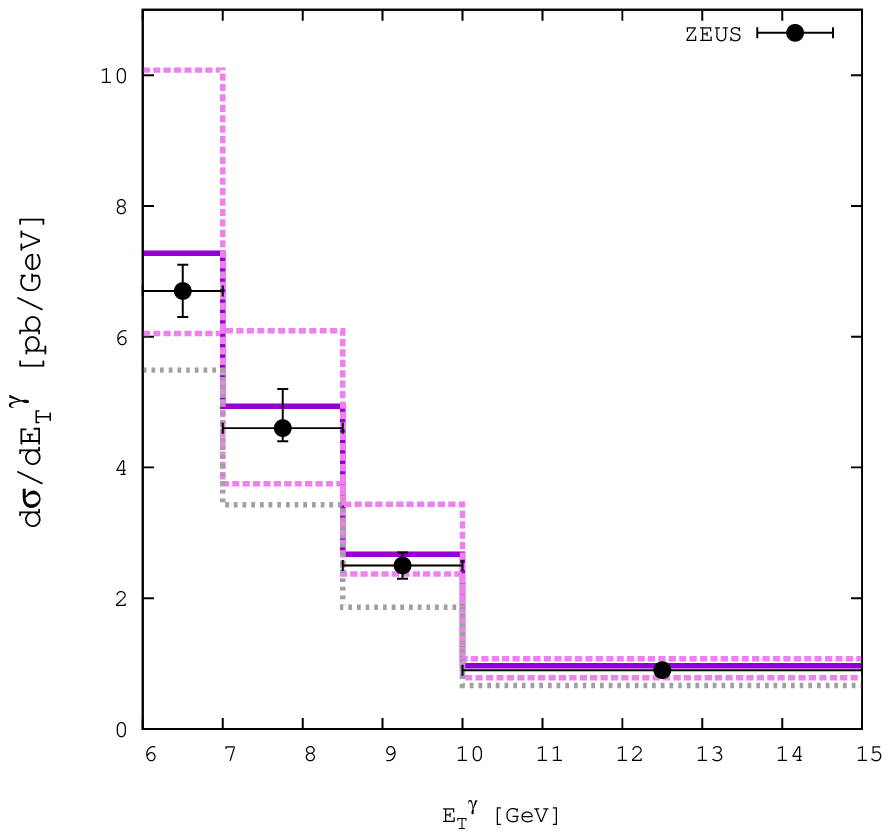}}
\subfloat[]
{\label{fig:image_22a2}
\includegraphics[width=0.5\textwidth]{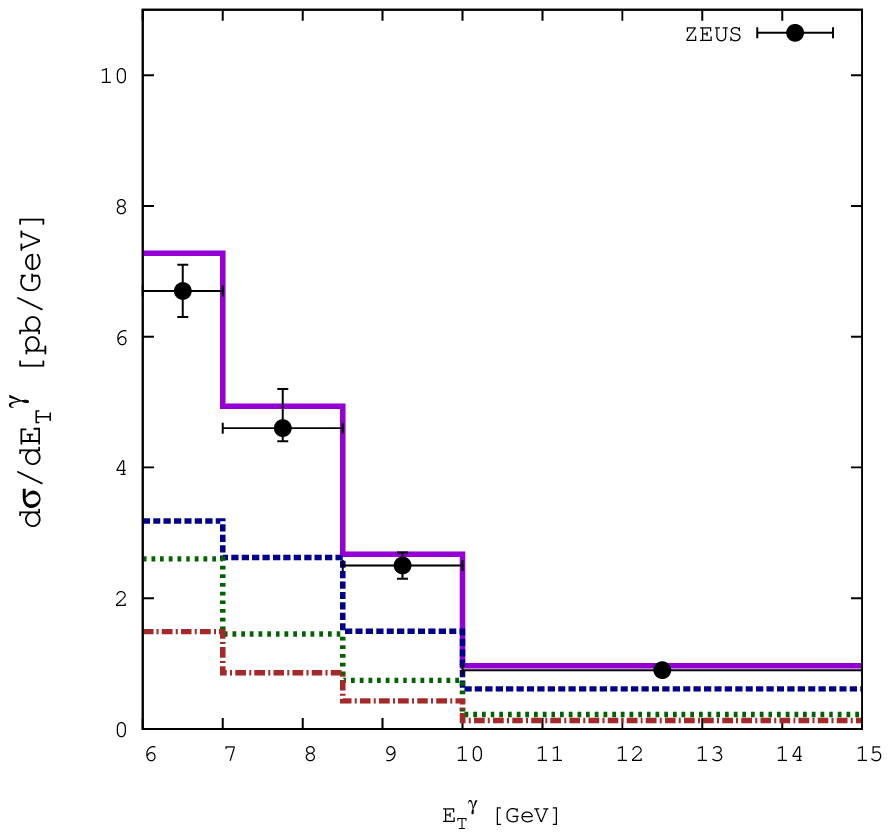}}

\subfloat[]%optionally add here a short text as a label]
{\label{fig:image_12b2}
\includegraphics[width=0.5\textwidth]{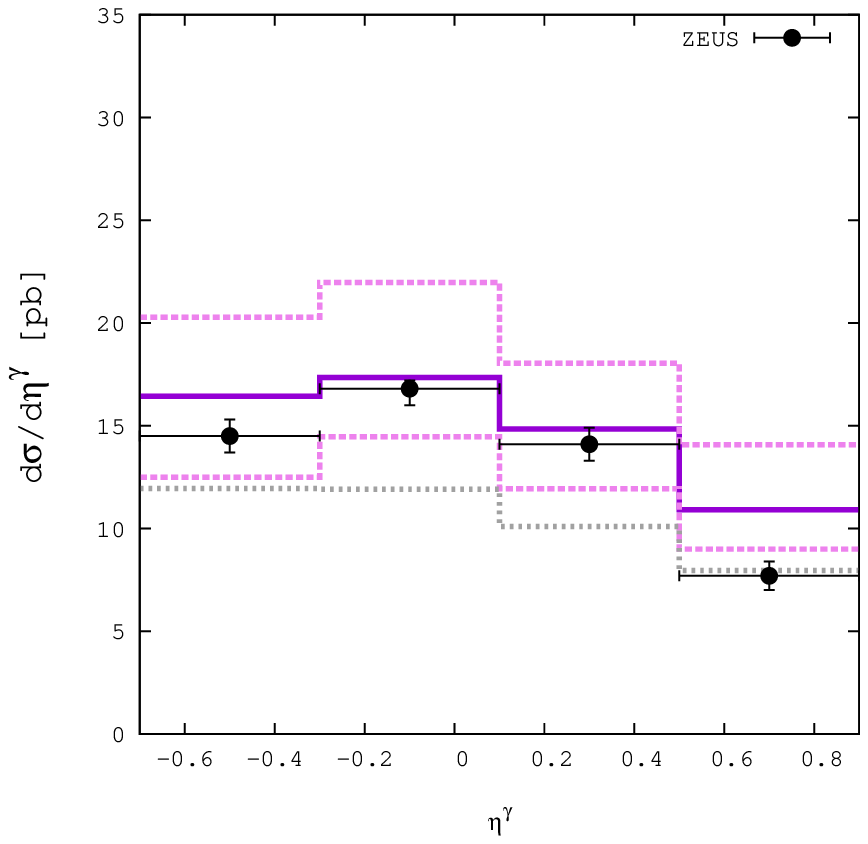}}
\subfloat[]
{\label{fig:image_22b2}
\includegraphics[width=0.5\textwidth]{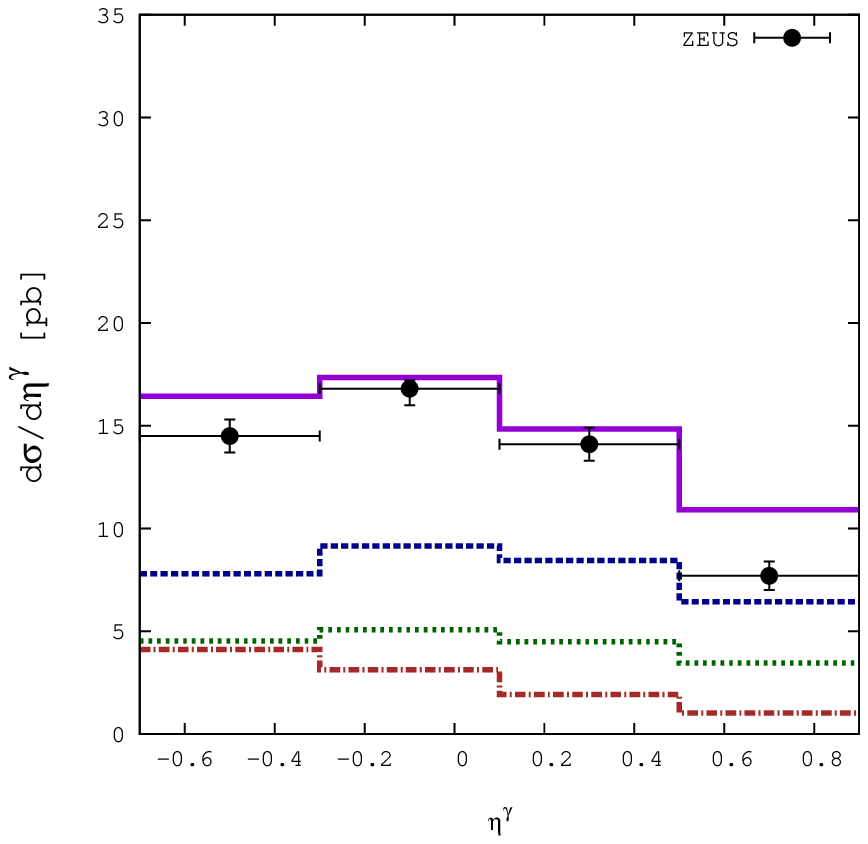}}

\centering
\caption{\label{ZEUS1} \it{The associated with a jet prompt photon photoproduction
cross section as a function of photon transverse energy $E_T^\gamma$ and
pseudo-rapidity $\eta^\gamma$ at HERA. The notations of the histograms are the same as in Fig.~\ref{inclusive-H1}. The experimental data are from ZEUS \cite{ZEUS-2013}.
}}
\end{figure}

\begin{figure}[t]
\centering
\subfloat[]%optionally add here a short text as a label]
{\label{fig:image_12c2}
\includegraphics[width=0.5\textwidth]{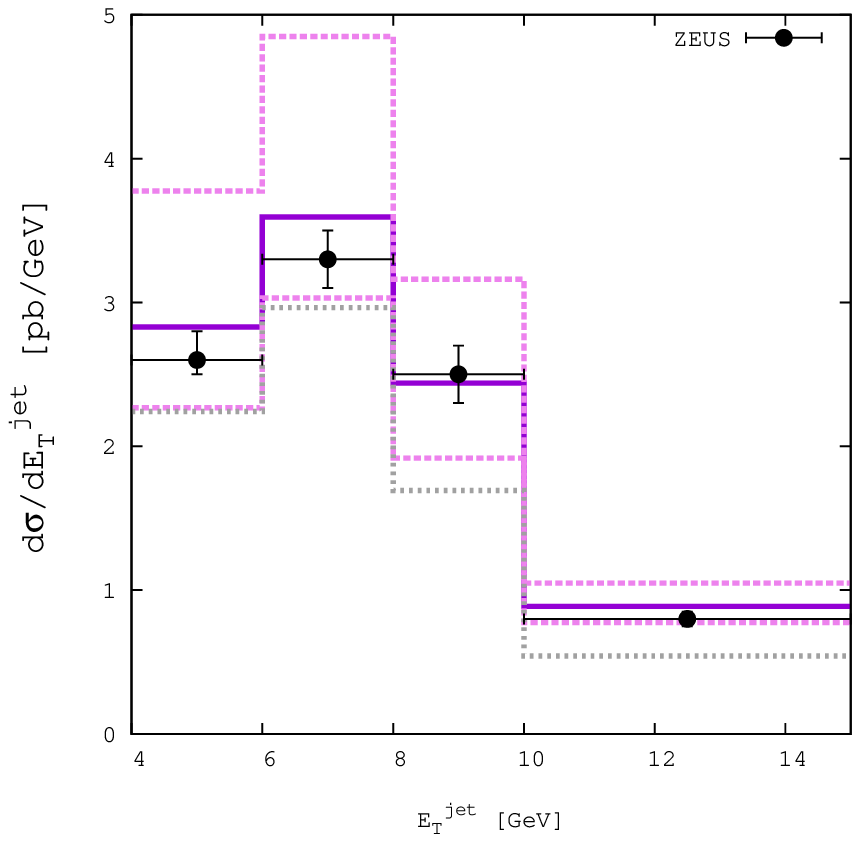}}
\subfloat[]
{\label{fig:image_22c2}
\includegraphics[width=0.5\textwidth]{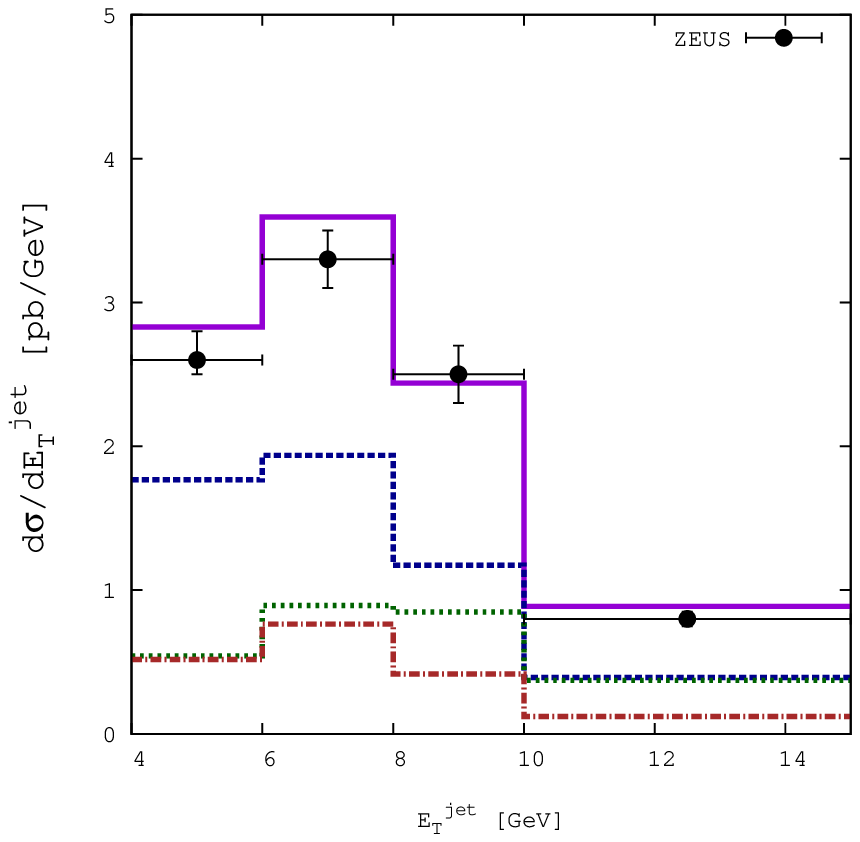}}

\subfloat[]{
\label{fig:image_32d2}
\includegraphics[width=0.5\textwidth]{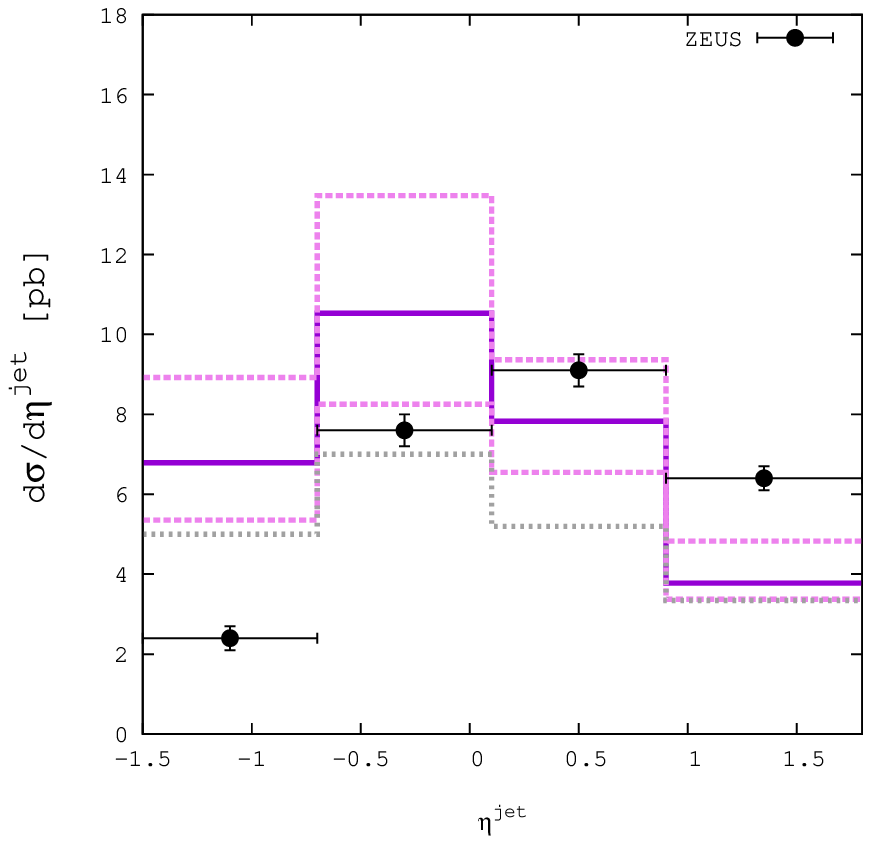}}
\subfloat[]{
\label{fig:image_42d2}
\includegraphics[width=0.5\textwidth]{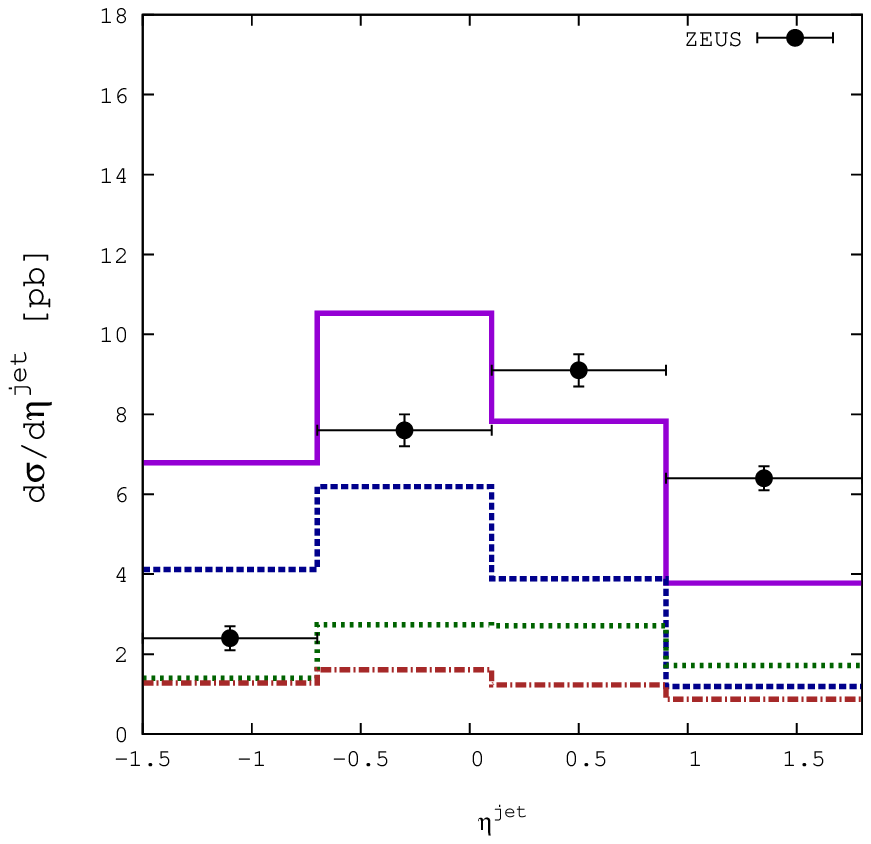}}

\centering
\caption{\label{ZEUS2} \it{The associated with a jet prompt photon photoproduction
cross section as a function of jet transverse energies $E_T^{jet}$ and
pseudo-rapidities $\eta^{jet}$ at HERA. The notations of the histograms are the same as in Fig.~\ref{inclusive-H1}. The experimental data are from ZEUS \cite{ZEUS-2013}.
}}
\end{figure}

Our numerical predictions are shown in Figs.~\ref{H11} --- \ref{ZEUS2} (on the left panels) 
in comparison with the H1 and ZEUS data \cite{H1-2009,ZEUS-2013}. The relative contributions of different subprocesses are shown in the right panels.
One can see that the situation is very similar to the inclusive production case.
The reasonably good description of the data for a number of measured distributions
is achieved except for the distributions on the 
$\eta^{\rm jet}$, where we see some disagreement in the shape. 
The same shape disagreement in the $\eta^{\rm jet}$ distributions is observed in the 
predictions based on the $2\to 2$ subprocesses (1) --- (4).
The possible reason of such discrepancy can be connected with the 
approximation for the jet determination which was described above and which was used in both type of calculations.
Note that the predictions based on the former scheme give the results tending to 
underestimate the data, while the approach based on the 
$2\to 3$ subprocesses (5) --- (7) shows a better agreement.
We find that the box contribution (7) is important in the photon and jet associated production
case also. One can see that its contribution is comparable with the $\gamma q\to\gamma q g$ subprocess. 

Other important variables in prompt photon photoproduction investigations are the 
longitudinal momenta of the partons in the colliding particles. The momentum fraction of the initial photon is introduced in the ZEUS analyses\cite{ZEUS-2013} as the following:
\begin{equation}
x_\gamma^{\rm obs}=\frac{E^\gamma_T e^{-\eta^\gamma}+E_T^{\rm jet}e^{-\eta^{\rm jet}}}{2yE_e}.
\end{equation}

\noindent At $x_\gamma^{\rm obs}>0.85$ the cross section is dominated by the direct initial photon contributions, whereas at lower $x_\gamma^{\rm obs}$ the resolved photon contribution dominate.
The H1 collaboration refers to $x_\gamma^{\rm LO}$ and $x_p^{\rm LO}$ observables\cite{H1-2009} given by:
\begin{equation}
x_\gamma^{\rm LO}=\frac{E_T^\gamma(e^{-\eta^\gamma}+e^{-\eta^{\rm jet}})}{2yE_e},\quad x_p^{\rm LO}=\frac{E_T^\gamma(e^{-\eta^\gamma}+e^{\eta^{\rm jet}})}{2E_p}.
\end{equation}

\begin{figure}
\centering
\subfloat[]{
\label{fig:image_32d2f}
\includegraphics[width=0.5\textwidth]{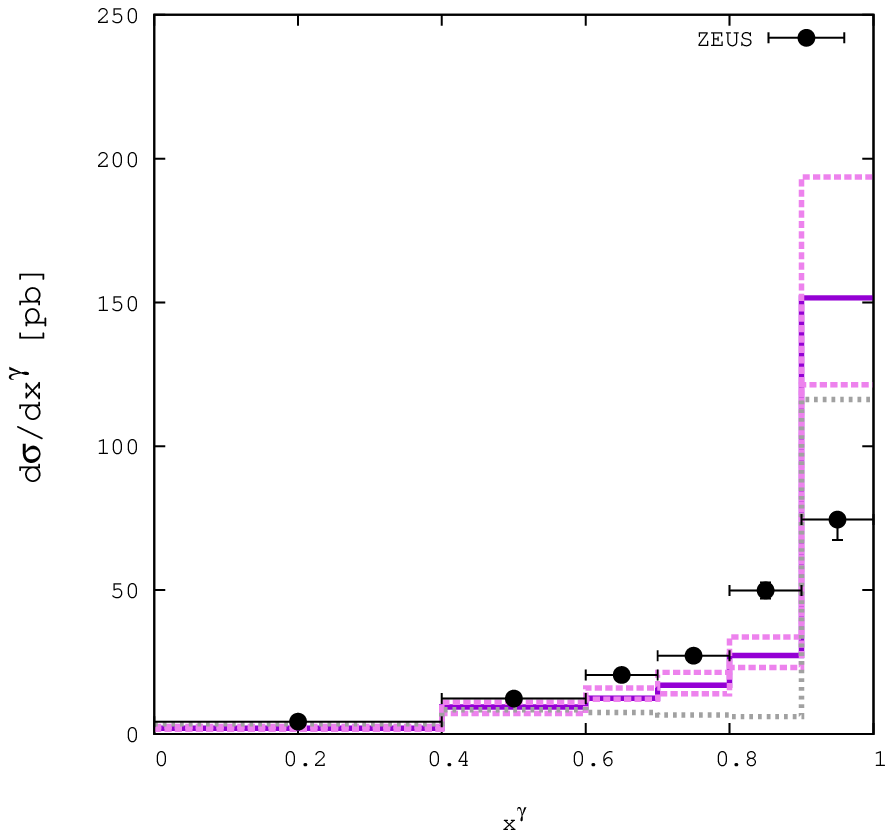}}
\subfloat[]{
\label{fig:image_42d2f}
\includegraphics[width=0.5\textwidth]{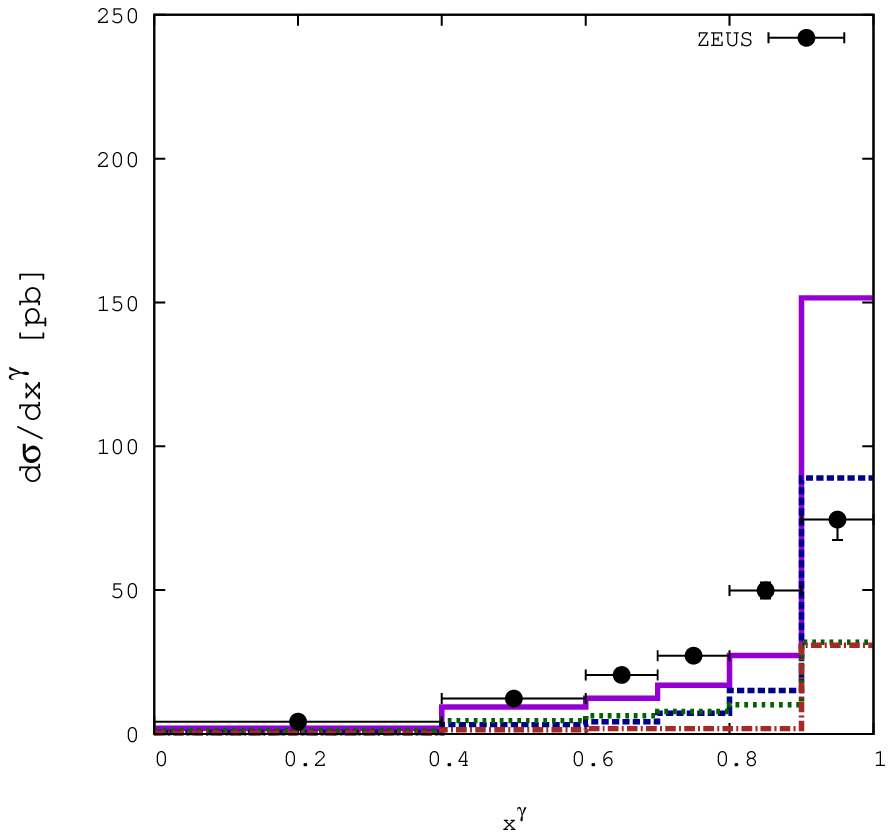}}

\centering
\caption{\label{x-ZEUS} \it{The associated with a jet prompt photon photoproduction
cross section as a function of the longitudinal momentum of a parton from the
initial photon $x_\gamma^{obs}$ at HERA The notations of the histograms are the same as in Fig.~\ref{inclusive-H1}. The experimental data are from ZEUS \cite{ZEUS-2013}.
}}
\end{figure}

\begin{figure}
\centering
\subfloat[]%optionally add here a short text as a label]
{\label{fig:image_16}
\includegraphics[width=0.5\textwidth]{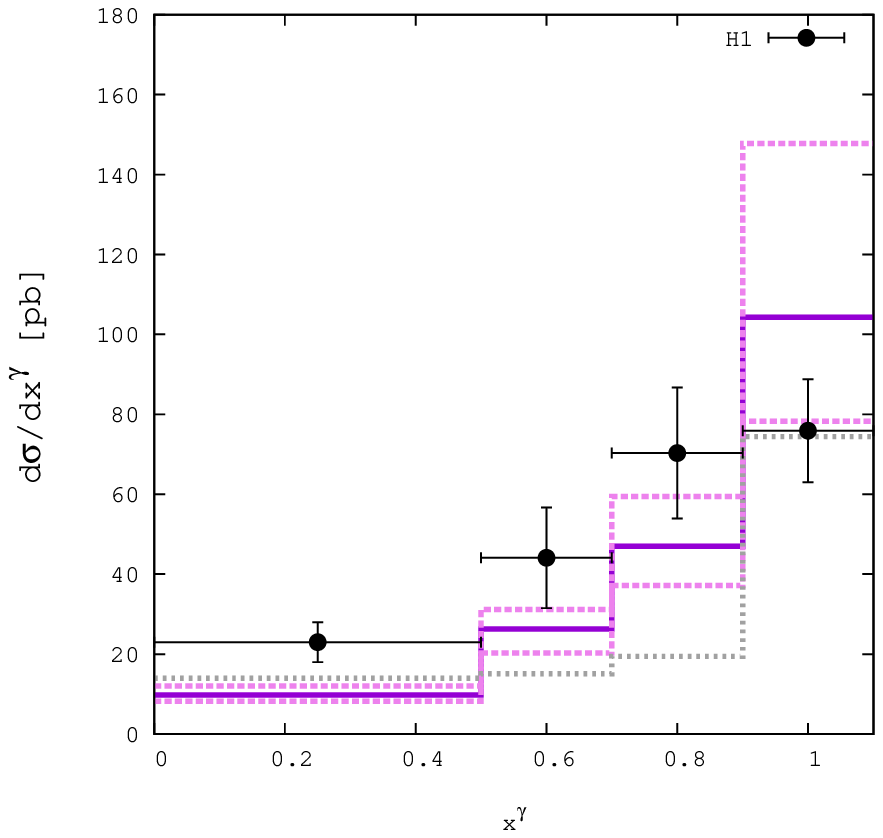}}
\subfloat[]
{\label{fig:image_26}
\includegraphics[width=0.5\textwidth]{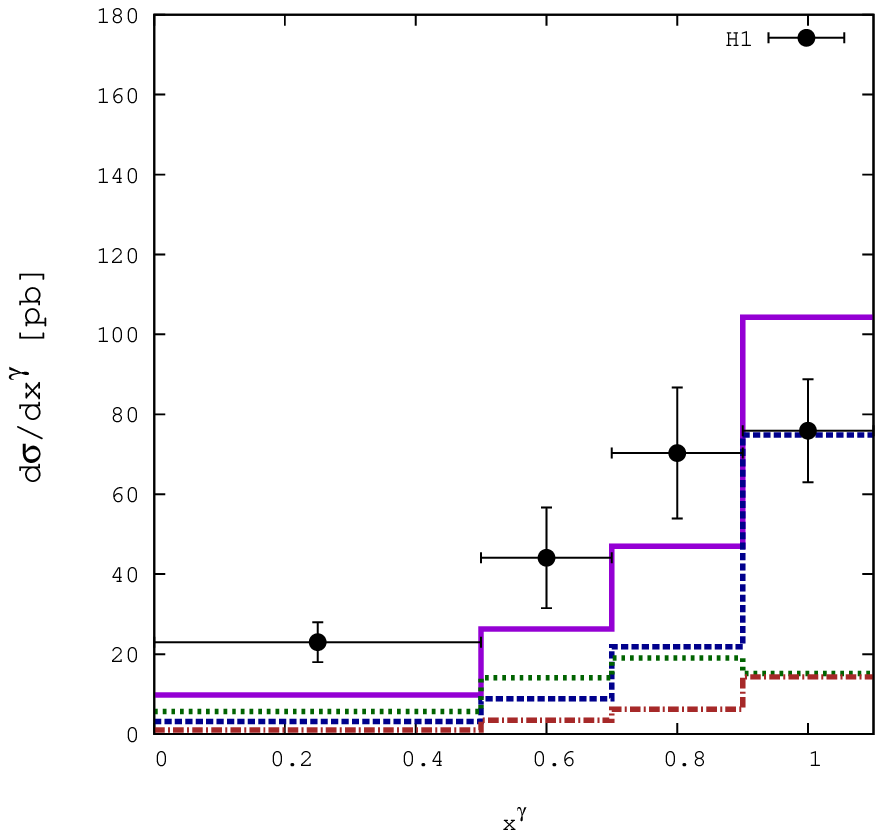}}

\subfloat[]%optionally add here a short text as a label]
{\label{fig:image_16g}
\includegraphics[width=0.5\textwidth]{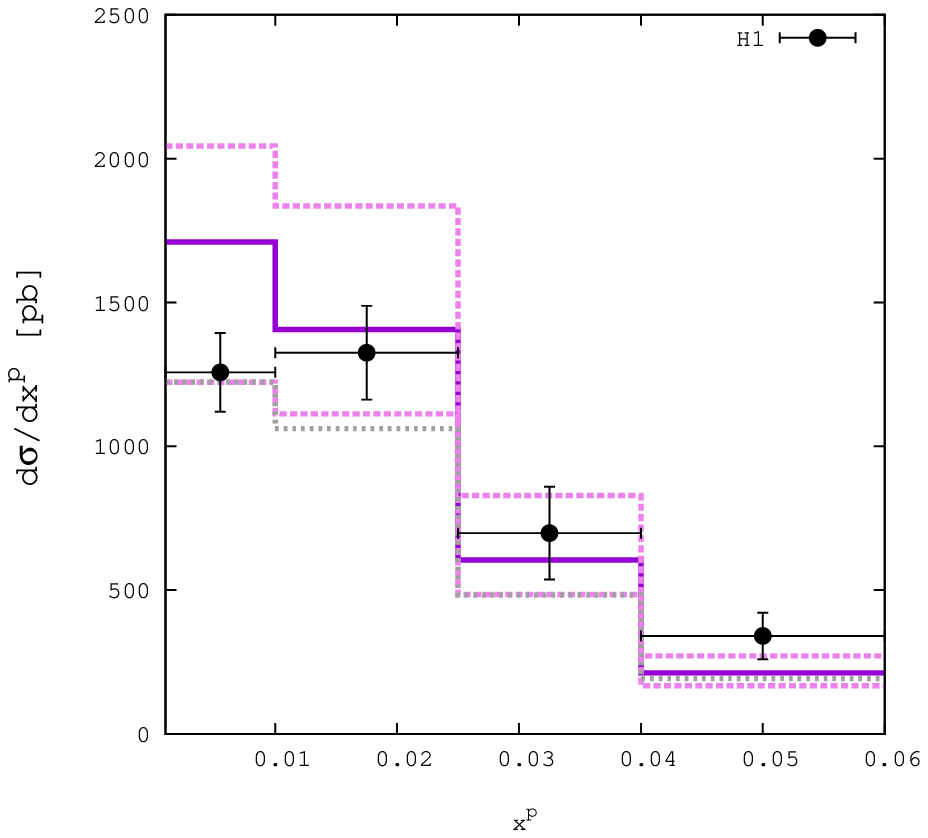}}
\subfloat[]
{\label{fig:image_26g}
\includegraphics[width=0.5\textwidth]{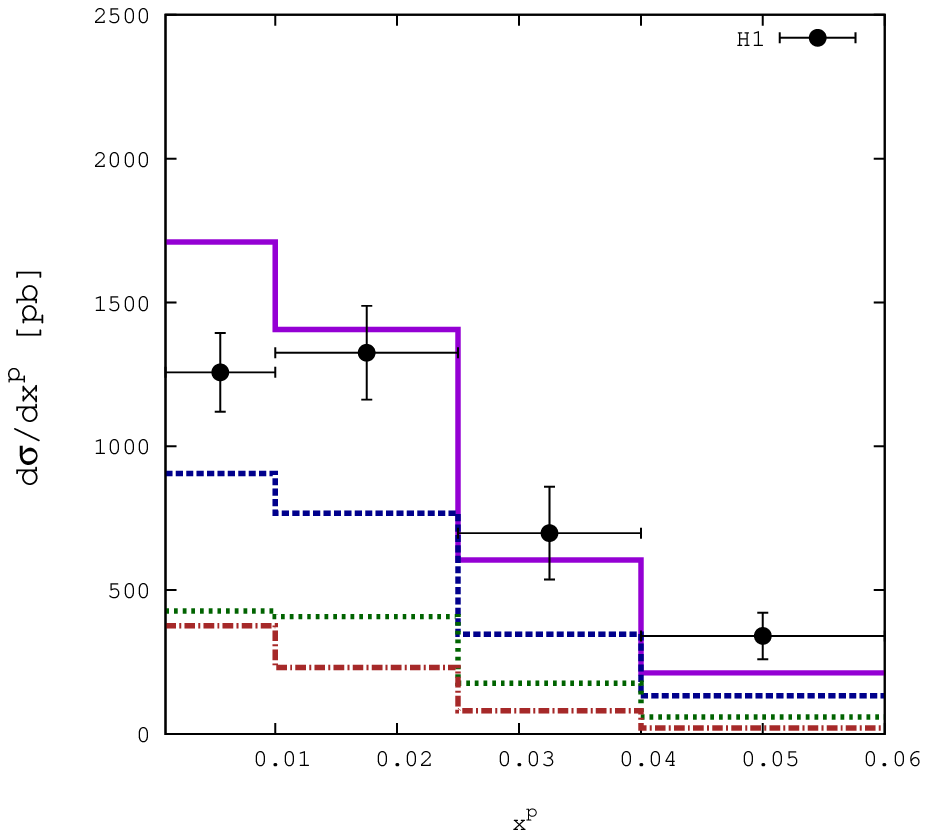}}

\centering
\caption{\label{x-H1} \it{The associated with a jet prompt photon photoproduction
cross section as a function of the $x_\gamma^{LO}$ and $x_p^{LO}$ variables at HERA.
The notations of the histograms are the same as in Fig.~\ref{inclusive-H1}. The experimental data
are from H1 \cite{H1-2009}.
}}
\end{figure}

\noindent
Our predictions for these observables 
are shown in Figs.~\ref{x-ZEUS} --- \ref{x-H1} in comparison with the H1 and ZEUS data.
We demonstrate that, in the framework of $k_T$-factorization, the 
subprocesses (\ref{Gq1}) and (\ref{Gg}) allow us to take into 
account both direct and resolved contributions without introducing any 
parton densities in a photon. One can see that the direct region with 
$x_\gamma^{\rm obs} > 0.85$ is dominated by the subprocess (\ref{Gq1}), which 
incorporates the contribution from the subprocess (\ref{Gq}). In the resolved photon region,
where $x_\gamma^{\rm obs}<0.85$, the contribution of the 
subprocess (\ref{Gg}) becomes more important, since it contains the single resolved 
photon contribution (\ref{qg}).
We point out that the approach based on the $2\to3$ subprocesses (5) --- (7) 
shows a better agreement with the data at intermediate $0.6<x_\gamma^{\rm obs}<0.9$ 
compared with the approach based on the $2\to 2$ subprocesses (1) --- (4).
This is a result of more accurate treatment of jet production kinematics 
achieved in a presented approach.

%Another set of data was presented by the H1 collaboration in \cite{H1-2009}. The restrictions on the prompt photon are the same as in the inclusive case. For the jets the kinematical region is $E_T^{jet}>4.5$ GeV and $-1.3<\eta^{jet}<2.3$. On Fig.~\ref{H1} the results of the theoretical calculations are presented. The notations are the same. One can see a good description in all graphs except for the $\eta^{jet}$ one, where a qualitative disagreement takes place. This can be connected with the approximate treatment of the evolution chain jets in our approach, since we don't use a full Monte-Carlo generator.

%Recently the ZEUS collaboration presented a new study of prompt photon associated with a jet photoproduction at HERA \cite{ZEUS-2013}. To the inclusive prompt photon restrictions the following jet limitations were added: $4<E_T^{jet}<35$ GeV and $-1.5<\eta^{jet}<1.8$. The comparison of this data with the $k_T$-factorization calculations are presented on the Fig.~\ref{ZEUS}. The notations are the same. As in the previous case one can see a reasonable description of all the data except for $\eta^{jet}$ one.

%Our prediction for $x_\gamma^{LO}$ and $x_p^{LO}$ distributions in the comparison with the H1 data \cite{H1-2009} are presented on the Fig.~\ref{x-H1}.

\section{Conclusion}

We have reconsidered the inclusive and associated jet prompt photon photoproduction 
at HERA in the framework of $k_T$-factorization QCD formalism. 
The proposed approach is based on the 
${\cal O}(\alpha^2\alpha_s)$ off-shell amplitudes 
for $\gamma q\to\gamma gq$ and $\gamma g^*\to\gamma q\bar q$ partonic subprocesses.
Similar consideration had a success in the description of the Tevatron data
on the associated photon and heavy quark jet taken by the D0 and CDF collaborations \cite{D0,CDF}.
We have taken into account also the ${\cal O}(\alpha^2\alpha_s^2)$ box 
contributions $\gamma g\to\gamma g$ to the production cross sections. 
The unintegrated (or transverse momentum dependent) parton densities in the proton have been 
determined using Kimber-Martin-Ryskin (KMR) prescription. 
We have demonstrated that the present approach results in a better agreement with the 
HERA data in contrast with the previous $k_T$-factorization predictions 
based on the $2\to 2$ subprocesses. We find that the box contribution is sizeable and amounts up to $\sim15\%$ of the 
calculated total cross section. The obtained results are important for the further 
theoretical and experimental investigations of prompt photon production 
associated with the hadronic jet(s) at the LHC energies.

\section*{Acknowledgements}
We would like to thank H. Jung for careful reading the manuscript and very useful remarks. We thank also P.\,Bussey, A.\,Iudin and I.\,Skillicorn for a useful discussion on the experimental data and the obtained results.
This research was supported in part by the FASI of Russian Federation
(grant NS-3920.2012.2), RFBR grants 12-02-31030 and 13-02-01060 and the grant of the Ministry of education and sciences
of Russia (agreement 8412).
A.L. and N.Z. are also grateful to DESY Directorate for the
support in the framework of Moscow---DESY project on Monte-Carlo implementation for
HERA---LHC.

%\newpage

\begin{figure}
\center
\begin{fmfgraph*}(120,80)
\fmfleftn{i}{2}\fmfrightn{o}{3}
\fmflabel{$g$}{i1}\fmflabel{$g$}{i2}
\fmflabel{$\bar q$}{o1}\fmflabel{$q$}{o3}\fmflabel{$\gamma$}{o2}
\fmf{gluon}{i1,v1}
\fmf{boson}{i2,v3}
\fmf{fermion}{o1,v1,v2,v3,o3}
\fmffreeze
\fmf{boson}{o2,v2}
\end{fmfgraph*}
%\begin{fmfgraph*}(30,40)
%\end{fmfgraph*}
\hspace*{1cm}
\begin{fmfgraph*}(120,80)
\fmfleftn{i}{2}\fmfrightn{o}{3}
\fmflabel{$g$}{i1}\fmflabel{$g$}{i2}
\fmflabel{$q$}{o1}\fmflabel{$\bar q$}{o3}\fmflabel{$\gamma$}{o2}
\fmf{gluon}{i1,v1}
\fmf{boson}{i2,v3}
\fmf{fermion}{o3,v3,v2,v1,o1}
\fmffreeze
\fmf{boson}{o2,v2}
\end{fmfgraph*}
%\begin{fmfgraph*}(30,40)
%\end{fmfgraph*}
\hspace*{1cm}
\begin{fmfgraph*}(120,80)
\fmfleftn{i}{2}\fmfrightn{o}{3}
\fmflabel{$g$}{i1}\fmflabel{$g$}{i2}
\fmflabel{$q$}{o1}\fmflabel{$\bar q$}{o3}\fmflabel{$\gamma$}{o2}
\fmf{gluon}{i1,v1}
\fmf{boson}{i2,v2}
\fmf{fermion}{o3,v3,v2,v1,o1}
%\fmffreeze
\fmf{boson}{o2,v3}
\end{fmfgraph*}
%\begin{fmfgraph*}(30,40)
%\end{fmfgraph*}
%\end{figure}
\newline
\newline
\newline
%\newline
%\newline
%\begin{figure}
%\begin{fmfgraph*}(20,40)
%\end{fmfgraph*}
\begin{fmfgraph*}(120,80)
\fmfleftn{i}{2}\fmfrightn{o}{3}
\fmflabel{$g$}{i1}\fmflabel{$g$}{i2}
\fmflabel{$\bar q$}{o1}\fmflabel{$q$}{o3}\fmflabel{$\gamma$}{o2}
\fmf{gluon}{i1,v2}
\fmf{boson}{i2,v3}
\fmf{fermion}{o1,v1,v2,v3,o3}
%\fmffreeze
\fmf{boson}{o2,v1}
\end{fmfgraph*}
%\begin{fmfgraph*}(30,40)
%\end{fmfgraph*}
\hspace*{1cm}
\begin{fmfgraph*}(120,80)
\fmfleftn{i}{2}\fmfrightn{o}{3}
\fmflabel{$g$}{i1}\fmflabel{$g$}{i2}
\fmflabel{$\bar q$}{o1}\fmflabel{$q$}{o3}\fmflabel{$\gamma$}{o2}
\fmf{gluon}{i1,v1}
\fmf{boson}{i2,v2}
\fmf{fermion}{o1,v1,v2,v3,o3}
%\fmffreeze
\fmf{boson}{o2,v3}
\end{fmfgraph*}
%\begin{fmfgraph*}(30,40)
%\end{fmfgraph*}
\hspace*{1cm}
\begin{fmfgraph*}(120,80)
\fmfleftn{i}{2}\fmfrightn{o}{3}
\fmflabel{$g$}{i1}\fmflabel{$g$}{i2}
\fmflabel{$q$}{o1}\fmflabel{$\bar q$}{o3}\fmflabel{$\gamma$}{o2}
\fmf{gluon}{i1,v2}
\fmf{boson}{i2,v3}
\fmf{fermion}{o3,v3,v2,v1,o1}
%\fmffreeze
\fmf{boson}{o2,v1}
\end{fmfgraph*}
%\begin{fmfgraph*}(30,40)
%\end{fmfgraph*}
%\begin{fmfgraph*}(30,40)
%\end{fmfgraph*}
\hspace*{1cm}
%\begin{fmfgraph*}(30,40)
%\end{fmfgraph*}
\newline
\newline
\newline
\begin{fmfgraph*}(120,80)
\fmfleftn{i}{2}\fmfrightn{o}{3}
\fmflabel{$g$}{i1}\fmflabel{$g$}{i2}
\fmflabel{$\bar q$}{o1}\fmflabel{$q$}{o3}\fmflabel{$\gamma$}{o2}
\fmf{boson}{i2,v1}
\fmf{boson}{o3,v2}
\fmf{fermion}{i1,v1,v2,v3,o1}
\fmffreeze
\fmf{gluon}{o2,v3}
\end{fmfgraph*}
%\begin{fmfgraph*}(30,40)
%\end{fmfgraph*}
\hspace*{1cm}
\begin{fmfgraph*}(120,80)
\fmfleftn{i}{2}\fmfrightn{o}{3}
\fmflabel{$g$}{i1}\fmflabel{$g$}{i2}
\fmflabel{$q$}{o1}\fmflabel{$\bar q$}{o3}\fmflabel{$\gamma$}{o2}
\fmf{boson}{i2,v1}
\fmf{gluon}{o3,v2}
\fmf{fermion}{i1,v1,v2,v3,o1}
\fmffreeze
\fmf{boson}{o2,v3}
\end{fmfgraph*}
%\begin{fmfgraph*}(30,40)
%\end{fmfgraph*}
\hspace*{1cm}
\begin{fmfgraph*}(120,80)
\fmfleftn{i}{2}\fmfrightn{o}{3}
\fmflabel{$g$}{i1}\fmflabel{$g$}{i2}
\fmflabel{$q$}{o1}\fmflabel{$\bar q$}{o3}\fmflabel{$\gamma$}{o2}
\fmf{boson}{i2,v2}
\fmf{gluon}{o1,v1}
\fmf{fermion}{i1,v1,v2,v3,o3}
%\fmffreeze
\fmf{boson}{o2,v3}
\end{fmfgraph*}
%\begin{fmfgraph*}(30,40)
%\end{fmfgraph*}
%\end{figure}
\newline
\newline
\newline
%\newline
%\newline
%\begin{figure}
%\begin{fmfgraph*}(20,40)
%\end{fmfgraph*}
\begin{fmfgraph*}(120,80)
\fmfleftn{i}{2}\fmfrightn{o}{3}
\fmflabel{$g$}{i1}\fmflabel{$g$}{i2}
\fmflabel{$\bar q$}{o1}\fmflabel{$q$}{o3}\fmflabel{$\gamma$}{o2}
\fmf{boson}{i2,v2}
\fmf{boson}{o1,v1}
\fmf{fermion}{i1,v1,v2,v3,o3}
%\fmffreeze
\fmf{gluon}{o2,v3}
\end{fmfgraph*}
%\begin{fmfgraph*}(30,40)
%\end{fmfgraph*}
\hspace*{1cm}
\begin{fmfgraph*}(120,80)
\fmfleftn{i}{2}\fmfrightn{o}{3}
\fmflabel{$g$}{i1}\fmflabel{$g$}{i2}
\fmflabel{$\bar q$}{o1}\fmflabel{$q$}{o3}\fmflabel{$\gamma$}{o2}
\fmf{boson}{i2,v3}
\fmf{boson}{o1,v1}
\fmf{fermion}{i1,v1,v2,v3,o3}
\fmffreeze
\fmf{gluon}{o2,v2}
\end{fmfgraph*}
%\begin{fmfgraph*}(30,40)
%\end{fmfgraph*}
\hspace*{1cm}
\begin{fmfgraph*}(120,80)
\fmfleftn{i}{2}\fmfrightn{o}{3}
\fmflabel{$g$}{i1}\fmflabel{$g$}{i2}
\fmflabel{$q$}{o1}\fmflabel{$\bar q$}{o3}\fmflabel{$\gamma$}{o2}
\fmf{boson}{i2,v3}
\fmf{gluon}{o1,v1}
\fmf{fermion}{i1,v1,v2,v3,o3}
\fmffreeze
\fmf{boson}{o2,v2}
\end{fmfgraph*}
%\begin{fmfgraph*}(30,40)
%\end{fmfgraph*}
%\begin{fmfgraph*}(30,40)
%\end{fmfgraph*}
\hspace*{1cm}
\begin{fmfgraph*}(30,40)
\end{fmfgraph*}
\caption{\it The $2\to3$ diagrams considered}
\label{2to3}
\end{figure}

\end{fmffile}

\begin{thebibliography}{99}
\bibitem{ZEUS-2000}
J.\,Breitweg et al. (ZEUS Collaboration), Phys.\,Lett. \textbf{B 472}, 175 (2000).
\bibitem{H1-2004}
A.\,Atkas et al. (H1 Collaboration), Eur.\,Phys.\,J. \textbf{C 38}, 437 (2005).
\bibitem{ZEUS-2006} 
S.\,Chekanov et al. (ZEUS Collaboration), Eur.\,Phys.\,J. \textbf{C 49}, 511 (2007).
\bibitem{H1-2009} 
F.D.\,Aaron et al. (H1 Collaboration), Eur.\,Phys.\,J. \textbf{C 66}, 17 (2010).
\bibitem{FGH} 
M.\,Fontannaz, J.P.\,Guillet, G.\,Heinrich, Eur.\,Phys.\,J. \textbf{C 21}, 303 (2001).
\bibitem{FH} 
M.\,Fontannaz, G.\,Heinrich, Eur.\,Phys.\,J. \textbf{C 34}, 191 (2004).
\bibitem{ZK} 
A.\,Zembrzuski, M.\,Krawczyk, hep-ph/0309308.
\bibitem{LZ-HERA1} 
A.V.\,Lipatov, N.P.\,Zotov, Phys.\,Rev. \textbf{D 72}, 054002 (2005).
\bibitem{LZ-HERA2}
A.V.\,Lipatov, N.P.\,Zotov, Phys.\,Rev. \textbf{D 81}, 094027 (2010).
\bibitem{kt1} V.N.\,Gribov, E.M.\,Levin, M.G.\,Ryskin, Phys.\,Rep. \textbf{100}, 1 (1983).
\bibitem{kt2} S.\,Catani, M.\,Ciafoloni, F.\,Hautmann, Nucl.\,Phys. \textbf{B 366}, 135 (1991).
\bibitem{kt3} J.C.\,Collins, R.K.\,Ellis, Nucl.\,Phys. \textbf{B 360}, 3 (1991).
\bibitem{kt4} E.M.\,Levin, M.G.\,Ryskin, Yu.M.\,Shabelsky, A.G.\,Shuvaev, Sov.\,J.\,Nucl.\,Phys. \textbf{53}, 657 (1991).
\bibitem{BFKL1} E.A.\,Kuraev, L.N.\,Lipatov, V.S.\,Fadin, Sov.\,Phys.\,JETP \textbf{44}, 443 (1976); \textbf{45}, 199 (1977).
\bibitem{BFKL2} I.I.\,Balitsky, L.N.\,Lipatov, Sov.\,J.\,Nucl.\,Phys. \textbf{28}, 822 (1978).
%\bibitem{CCFM1} M.\,Ciafaloni, Nucl.\,Phys. \textbf{B 296}, 49 (1988).
%\bibitem{CCFM2} S.\,Catani, F.\,Fiorani, G.\,Marchesini, Phys.\,Lett. \textbf{B 234}, 339 (1990); Nucl. Phys. \textbf{B~336}, 18 (1990).
%\bibitem{CCFM3} G.\,Marchesini, Nucl.\,Phys. \textbf{B 445}, 49 (1995).
\bibitem{small-x}
B.\,Andersson et al. (Small-$x$ Collaboration), Eur.\,Phys.\,J. \textbf{C 25}, 77 (2002); J.\,Andersen et al. (Small-$x$ Collaboration), Eur.\,Phys.\,J. \textbf{C 35}, 67 (2004); Eur.\,Phys.\,J. \textbf{C 48}, 67 (2006).
\bibitem{ZEUS-2013} 
A.\,Iudin et al. (ZEUS Collaboration), talk given at DIS'2013.
%\bibitem{GribovLipatov}
%V.N.\,Gribov and L.N.\,Lipatov, Sov.\,Journ.\,Nucl.\,Phys. \textbf{15}, 438, 675 (1972).
%\bibitem{AltarelliParisi}
%G.\,Altarelli, G.\,Parisi, Nucl.\,Phys. \textbf{126}, 298 (1977).
%\bibitem{Dokshitzer}
%Yu.L.\,Dokshitzer, Sov.\,Phys.\,JETP \textbf{46}, 641 (1977).
%\bibitem{LZ-PP-2007}
%A.V.\,Lipatov, N.P.\,Zotov, J.\,Phys. \textbf{G 34}, 219 (2007).
%\bibitem{BLZ-PP-2008}
%S.P.\,Baranov, A.V.\,Lipatov, N.P.\,Zotov, Phys.\,Rev. \textbf{D 77}, 074024 (2008).
%\bibitem{PP} 
%A.V.\,Lipatov, M.A.\,Malyshev, N.P.\,Zotov, Phys.\,Lett. \textbf{B 699}, 93 (2011).
%\bibitem{BLZ-PP+b-2008}
%S.P.\,Baranov, A.V.\,Lipatov, N.P.\,Zotov, Eur.\,Phys.\,J. \textbf{C 56}, 371 (2008).
\bibitem{gamma+b} 
A.V.\,Lipatov, M.A.\,Malyshev, N.P.\,Zotov, JHEP \textbf{1205}, 104 (2012).
\bibitem{D0}
V.M.\,Abazov et al. (D0 Collaboration), Phys.\,Lett. \textbf{B 714}, 32 (2012); \textbf{B 719}, 354 (2013).
\bibitem{CDF}
T.\,Aaltonen et al. (CDF Collaboration), arXiv:1303.6136[hep-ex].
\bibitem{gamma+b-NLO}
T.P.\,Stavreva, J.F.\,Owens, Phys.\,Rev. \textbf{D 79}, 054017 (2009).
%\bibitem{LZ-hera2} 
%A.V.\,Lipatov, N.P.\,Zotov, Phys.\,Rev. \textbf{D 81}, 094027 (2010).
%\bibitem{kt2} S.\,Catani, M.\,Ciafoloni, F.\,Hautmann, Nucl.\,Phys. \textbf{B 366}, 135 (1991).
%\bibitem{kt3} J.C.\,Collins, R.K.\,Ellis, Nucl.\,Phys. \textbf{B 360}, 3 (1991).
%\bibitem{kt4} E.M.\,Levin, M.G.\,Ryskin, Yu.M.\,Shabelsky, A.G.\,Shuvaev, Sov.\,J.\,Nucl.\,Phys. \textbf{53}, 657 (1991).
%\bibitem{BFKL1} E.A.\,Kuraev, L.N.\,Lipatov, V.S.\,Fadin, Sov.\,Phys.\,JETP \textbf{44}, 443 (1976); \textbf{45}, 199 (1977).
\bibitem{KMR1}
M.A.\,Kimber, A.D.\,Martin, M.G.\,Ryskin,  Phys.\,Rev. \textbf{D 63}, 114027 (2001).
\bibitem{KMR2}
G.\,Watt, A.D.\,Martin, M.G.\,Ryskin, Eur.\,Phys.\,J. \textbf{C 31}, 73 (2003).
\bibitem{FORM}
J.A.M.\,Vermaseren, NIKHEF-00-023 (2000).
\bibitem{box}
E.L.\,Berger, E.\,Braaten, R.D.\,Field, Nucl.\,Phys. \textbf{B 239}, 52 (1984).
\bibitem{diphoton}
A.V.\,Lipatov, JHEP \textbf{1302}, 009 (2013).
\bibitem{MSTW}
A.D.\,Martin, W.J.\,Stirling, R.S.\,Thorne, G.\,Watt, Eur.\,Phys.\,J. \textbf{C 63}, 189 (2009).
\bibitem{VEGAS}
G.P. Lepage, J. Comput. Phys. \textbf{27}, 192 (1978).
%\bibitem{BaranovZotov} 
%S.P.\,Baranov, N.P.\,Zotov, Phys.\,Lett. \textbf{B 491}, 111 (2000).
\end{thebibliography}
\end{document}